\documentclass[useAMS,usenatbib]{mn2e}

\usepackage{graphicx}
\usepackage{amsmath}
\usepackage{amssymb}
\usepackage{natbib}
\usepackage{fixltx2e}

\usepackage{color}
\def\apjl{ApJL}
\def\apj{ApJ}
\def\aap{A\&A}
\def\araa{ARA\&A}
\def\aj{AJ}
\def\mnras{MNRAS}
\def\actaa{Acta Astronomica}
\def\nat{Nature}

\def\srcs{QSO B0218+357}
\def\srclens{[PBK93]\,B0218+357\,G}

\newcommand{\kom}[1]{\textcolor{black}{#1}}

\title[]
{Variability of GeV gamma-ray emission in \srcs\ due to microlensing on intermediate size structures}

\author[J. Sitarek \& W. Bednarek]{J. Sitarek \& W. Bednarek\\
Department of Astrophysics, University of \L\'od\'z, PL-90236 \L \'od\'z, Poland; jsitarek@uni.lodz.pl, bednar@uni.lodz.pl}

\begin{document}

\date{Accepted . Received ; in original form }

\pagerange{\pageref{firstpage}--\pageref{lastpage}} \pubyear{2010}

\maketitle

\label{firstpage}

\begin{abstract}
Strong gravitational lensing leads to an occurrence of multiple images, with different magnifications, of a lensed source.
Those magnifications can in turn be modified by microlensing on smaller mass scales within the lens. 
Recently, measurements of the changes in the magnification ratio of the individual images have been proposed as a powerful tool for estimation of the size and velocity of the emission region in the lensed source. 
The changes of the magnification ratios in blazars PKS1830-211 and \srcs , if interpreted as caused by a microlensing on individual stars, put strong constraints on those two variables. 
These constraints are difficult to accommodate with the current models of gamma-ray emission in blazars.
In this paper we study if similar changes in the magnification ratio can be caused by microlensing on intermediate size structures in the lensing galaxy.
We investigate in details three classes of possible lenses: globular clusters (GC), open clusters (OC) and giant molecular clouds (GMC).
We apply this scenario to the case of \srcs .
Our numerical simulations show that changes in magnifications with similar time scales can be obtained for relativistically moving emission regions with sizes up to 0.01\,pc in the case of microlensing on the cores of GCs or clumps in GMCs. 
From the density of such structures in spiral galaxies we estimate however that lensing in giant molecular clouds would be more common. 
\end{abstract}

\section{Introduction}

If a large mass, such as a galaxy, is located between a source of radiation and observer, it will bend the trajectories of the photons and distort the observed image. 
In particular it can lead to the occurrence of multiple images of the same source.  
In such cases, the radiation that is emitted at the same time from the source will travel along different paths. 
As a consequence, in the case of time variable sources, the observer will record the same variability pattern from the various images, but with a time delay dependent on the geometry of the source-lens-observer system.
This effect due to the mass distribution of the whole galaxy, is referred to as macrolensing, and may be accompanied by additional effects due to individual stars in the lensing galaxy\kom{,} i.e. microlensing. 
The latter affects the images on much smaller angular scales, not observable with imaging instruments.

However, the deflection of the photon's trajectories results in changes of the magnification that can be observed (see e.g. \citealp{wa06}).
The microlensing is sensitive to small changes in the size and the location of the emission region.
Thus, it can be used to find and study the morphology of sources well beyond the reach of the angular resolution of even radio instruments.
This technique is suitable for search of e.g. dark matter clumps (see e.g. \citealp{pa86}) and extrasolar planets (see e.g. \citealp{ud02}).
Recently, microlensing was used to explain the time variability properties displayed in the hundred MeV-GeV band by the known gravitationally-lensed blazars PKS1830-211 \citep{ab15,nvm15} and \srcs\ \citep{vn15}.
These authors show that the changes in the magnification ratio of the leading and trailing component are consistent with microlensing due to individual stars in the lensing galaxy, as long as the emission region is relatively small, $\sim 10^{14}-10^{15}$\,cm, and the relative speed of the source and the microlens is of the order of $10^3$\,km/s.
Those results are however at odds with the standard paradigm of blazars, where the high-energy emission is generated in compact regions moving with relativistic velocities along the jet. 
Relativistic velocities are needed to explain the observed properties of blazars, such as high luminosity during flares, fast intrinsic variability, and indirectly also the lack of strong absorption of TeV gamma rays.

In this paper we investigate whether the observed changes in the relative gamma-ray magnification of both components can be explained by microlensing on larger structures than stars.
The size of those objects would result in much larger regions in the source plane being magnified by a single microlensing event, than for the case of microlensing on individual stars.
Therefore, microlensing on such intermediate size objects, if plausible, can relax the strong constraints on the size of the gamma-ray emission region and its velocity. 
We study \kom{the} following classes of objects acting as possible lenses for such a process: OCs, GCs and GMCs.
We focus on the interpretation of the changes in the magnification pattern of the 2012 high state of \srcs .
In Section~\ref{sec:src} we introduce the blazar \srcs .
In Section~\ref{sec:microlens} we estimate how probable the microlensing is on various intermediate scale structures.
Using the inverted ray shooting method we compute the magnification maps for typical parameters of such structures in Section~\ref{sec:magn}. 
In Section~\ref{sec:conc} we discuss the plausibility of such a scenario to occur in \srcs\ and how it would affect the future observations of this source.

\section{\srcs}\label{sec:src}
\srcs\ is a blazar located at a redshift of $0.944$ \citep{li12}.
It is gravitationally lensed by \srclens \footnote{https://www.cfa.harvard.edu/castles/} located at a redshift of $0.68$ \citep{br93}.
Using $H_0=69.6$, $\Omega_M=0.286$ and $\Omega_{vac}=0.714$, those redshifts correspond to an angular distance to the source $D_s=1650$\,Mpc, to the lens $D_L=1480$\,Mpc and between the two: $D_{LS}=370$\,Mpc. 
The radio image shows two distinct components A and B with an angular separation of only 335\,mas and an Einstein's ring of a similar size \citep{od92}.
The two radio components are separated by a time delay of $10.5\pm 0.4$ days (B lagging behind A) and show a flux ratio of the A and B components of $\mu_A/\mu_B \approx 3.6$ \citep{bi99}.
\citet{co00} and \citet{em11}, using radio measurements over the same epoch, obtained similar values for the time delay, but with larger confidence intervals, $10.1^{+1.5}_{-1.6}$ days and $11.8\pm2.3$ days, respectively.
Those results are consistent with the earlier measurement reported by \citet{co96} of $12\pm3$ days.
The ratio of magnification fluxes varies from 3.7 to 2.6 over the frequency range 15.35 GHz to 1.65 GHz \citep{mi06}, presumably due to free-free absorption \citep{mi07}.

\srclens\ is most probably a spiral galaxy seen face-on, with spiral arms spreading out to a distance of $\sim$5\,kpc \citep{yo05}. 
Using the position of the galaxy obtained by \cite{yo05}, the offset of the A and B images to the centre of the lens (measured in its frame of reference) is 1.8\,kpc and 0.47\,kpc, respectively. 

In 2012 \srcs\ underwent a high state in gamma rays.
A series of outbursts was registered by the \kom{\textit{Fermi} Large Area Telescope (LAT) in the MeV-GeV} range \citep{ch14}.
\citet{vn15} claimed to be able to decompose the emission into two separated components, delayed by $\sim 11.5$\,days, showing changes in their magnification ratio by a factor of a few.
Interestingly, another flare of \srcs\ in 2014 allowed the detection of this source also in VHE gamma-rays \citep{atel6349, si15}.
\section{Microlensing on intermediate size structures}\label{sec:microlens}
In this section we consider lensing of \srcs\ by different types of intermediate size and mass structures (with masses $10^2-10^7M_\odot$). 
We investigate in details 3 object classes: GCs, OCs and GMCs. 
Such masses for the geometry of the observer-lens-source of \srcs\ would still result in microlensing effect as the separation of the individual images is too small.
Another possible target for microlensing would be dark matter substructures (see e.g. \citealp{mo99}). 
However, as the details of the dark matter distribution are not known and only estimated from the simulations, we do not consider lensing on dark matter in this work.

The Einstein radius, which determines the angular scale at which lensing can occur may be computed for a point like object with mass $M$ as: 
\begin{equation}
\theta_E (M)=\sqrt{\frac{4GM}{c^2}\times \frac{D_{LS}}{D_L D_S}}=
3.1\times10^{-8} \left(\frac{M}{10^4M_\odot}\right)^{1/2}\,[^\circ].\label{eq:re}
\end{equation}
Lensing can occur on the whole structure if, for the geometry of the \srcs\ source-lens system, $\theta_E$ is larger than the angular size of the lens.
However, even if $\theta_E$ is a factor of a few smaller than the size of the lens, the lensing might still occur on substructures, or the inner part of the lens. 
Finally, for a coherent lensing of the whole emission region to occur, $\theta_E$ must by larger than the size of the lensed source. 

\subsection{GCs and OCs}\label{sec:gc-oc} 
GCs are spherical, tightly bound by gravity, collections of stars.
They are normally composed of a few times $10^5$ late-type stars, within a typical half-mass radius of the order of a few pc.
They are distributed in the spherical galactic halo. 
In the Milky Way galaxy so far over 150 GC have been detected \citep{ha96}, but for example in the Andromeda galaxy $\sim 500$ are expected \citep{bh01}.
Giant elliptical galaxies, such as M87, can have as much as 13000 GCs \citep{mhh94}. 
The density of the stars at the radius R from the centre of the cluster can be described by the following profile (see \citealp{mi63, kp06}):
\begin{eqnarray}
D(R) = \left\{ \begin {array}{ll}
1,                      & R < R_{\rm c} \\
(R_{\rm c}/R)^2,          & R_{\rm c} < R < R_{\rm h} \\
(R_{\rm c}R_{\rm h})^2/R^4, & R_{\rm h} < R < R_{\rm t}, 
\end{array} \right.
\label{eqgc}
\end{eqnarray}
where $R_{\rm c}$, $R_{\rm t}$ and $R_{\rm h} = \sqrt{2R_{\rm c}R_{\rm t}/3}$ are the GC core, tidal and half-mass radii, respectively.

On the other hand, OCs are groups of up to a few thousand loosely bound stars within a few pc distance. 
They are common in spiral and irregular galaxies. 
In our galaxy over one thousand OCs have been detected (see e.g. \citealp{kh13}). 
It is expected however that the total number can be even 10 times larger. 
Contrary to GCs, they mostly populate the galactic plane. 
In both, GC and OC, radiation pressure and supernova explosions drive the gas away from the cluster. 
Emission of sources with sizes up to $\sim \theta_E D_S=0.9 \sqrt{M/10^4M_\odot}$\, pc can be magnified via microlensing on a GC or an OC. 

\begin{figure}
\includegraphics[width=0.49\textwidth]{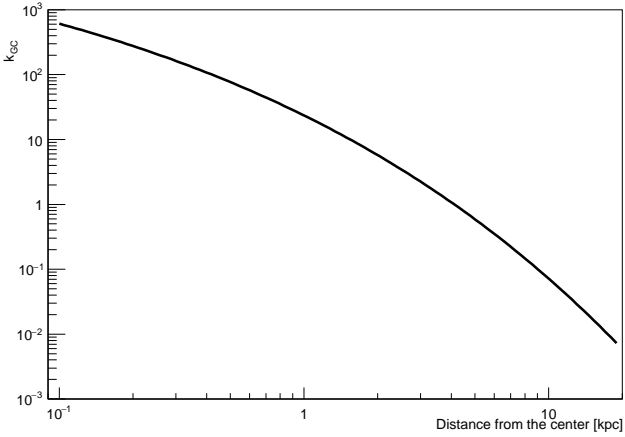}
\caption{Enhancement of the probability of lensing on a GC as a function of the distance of the image from the centre of the lensing galaxy for \srcs .
}\label{fig:kinhom}
\end{figure}

Let us now estimate how probable a lensing event is by computing the microlensing optical depth $\tau$. 
It can be estimated as $\tau=N k \pi (\theta_E(M) D_L)^2 / S$, where $N$ is the total number of GCs or OCs in the lensing galaxy and $S$ is the projected size of the region in which they are distributed.
$k$, dependent on the distance from the centre of the lensing galaxy, is the correction factor for the inhomogeneity of the surface density of the clusters.  
In the case of GCs, we can take $S_{GC} = \pi r^2_{halo}$, where $r_{halo}$ is the size of the halo in which they are distributed. 
We obtain:
\begin{equation}
\tau_{GC}=9.5\times 10^{-4} \frac{k_{\rm GC}}{30} \frac{M_{GC}}{10^5M_\odot} \frac{N_{GC}}{500} \left(\frac{r_{halo}}{10 \mathrm{kpc}}\right)^{-2}.
\label{eq:taugc}
\end{equation}
Note that this estimation does not depend on the orientation of the lensing galaxy. 
As the special density of GCs is strongly peaked towards the centre of a galaxy ($\propto R^{-3}$, see \citealp{hr79}) and the surface density is additionally enhanced by the projection effect, $k_{\rm GC}$ can obtain high values. 
In order to estimate it we use the \citet{va77} relation on the surface density of GCs in the Milky Way: $\log \sigma(R)=3.23-2.57(R/\mathrm{kpc})^{1/4}$. 
In order to use the above equation for \srclens , one has to correct the distance scale of $R$ by a factor of $2.5$ to take into account that \srclens\ is more compact than the Milky Way.
We then compute $k_{\rm GC}$ as the ratio of the surface density at the distance $R$ from the centre of the lensing galaxy, to a flat surface density up to a radius of $r_{halo}=10 \mathrm{kpc}$ and plot it in Fig.~\ref{fig:kinhom}.
For the distances of interest in \srcs\ we obtain $k_{\rm GC}(0.47\mathrm{\,kpc})\sim 80$, $k_{\rm GC}(1.8\mathrm{\,kpc})\sim 7$.

Similarly we can compute the optical depth for an OC seen face-on, assuming $S_{OC} = \pi r^2_{\rm disk}$, where $r_{\rm disk}$ is the radius of the galactic disk where the OCs are present,

\begin{equation}
\begin{split}
\tau_{OC, face}=&6.3\times 10^{-5} k_{\rm OC, face} \times \\
&\frac{M_{OC}}{10^4M_\odot} \frac{N_{OC}}{10000} \left(\frac{r_{\rm disk}}{10 \mathrm{kpc}}\right)^{-2}.
\end{split}
\end{equation}

However, if the same lens is seen edge-on the area covered by OCs is much smaller. 
It can be computed as $S_{OC}=4 r_{\rm disk} h_{\rm disk}$, with $h_{\rm disk}$ being the maximum height of the disk up to which the OCs are observed,

\begin{equation}
\begin{split}
\tau_{OC, edge}=&8.3\times 10^{-3} k_{\rm OC, edge} \times \\
&\frac{M_{OC}}{10^4M_\odot} \frac{N_{OC}}{10000} \left(\frac{r_{\rm disk}}{10 \mathrm{kpc}}\frac{h_{\rm disk}}{60 \mathrm{pc}}\right)^{-1}.
\end{split}
\end{equation}
Due to the projection effect, the lensing on OCs would be much more probable if the lensing galaxy itself is seen edge-on.
Note however, that in the case of \srcs , the optical image with the individual spiral arms is visible face-on.
The optical depth values are rather low making the lensing on GCs and OCs not very probable except for galaxies with a greater abundance of star clusters.

\subsection{GMCs}
GMCs are large gas structures where intense star formation occurs.
They typically have masses of the order of $10^5 M_\odot$ and radii $\sim 20$\,pc \citep{bl93}. 
Nevertheless, there is a large spread in both of those parameters (see e.g. \citealp{mu11}).
GMC have complicated structures, with filaments, clumps and cores (see \citealp{wbm00} and references within).
There are over $10^4$ GMCs in the Milky Way, with $10^3$ of them having masses above $2\times10^5 M_\odot$ \citep{mu11}. 

Since GMCs are much more irregular than star clusters, microlensing can occur on smaller scales (clumps) within them.
We can compute roughly the optical depth for a source being microlensed at a given moment by a clump in a GMC:
\begin{equation}\label{eq6}
\begin{split}
\tau_{\rm clump}&=k_{\rm GMC} N_{\rm GMC} N_{\rm clump} (\theta_E(M_{\rm clump}) D_L)^2 / r_{\rm disk}^2\\
&=2 \times 10^{-3} \frac{k_{\rm GMC}}{3} \frac{M_{\rm GMC}}{10^5M_\odot} \frac{N_{\rm GMC}}{10^4} \left(\frac{r_{\rm disk}}{10 \mathrm{kpc}}\right)^{-2},
\end{split}
\end{equation}
where $N_{\rm clump}$ and $M_{\rm clump}$ are the number and mass of clumps in a typical GMC and $M_{\rm GMC}=N_{\rm clump}M_{\rm clump}$.
Based on the H$_2$ distribution in the Galaxy, estimated with CO measurements \citep{sa84}, and rescaling them for the different sizes of \srclens\ and the Galaxy,  the inhomogeneity factor $k_{\rm GMC}$ is expected to be of the order of 1 for the A image, and of the order of 10 for the B image.
Moreover, as the projected position of the lensed source traverses the GMC, it can cross multiple individual clumps on time scales of months.
Therefore, the probability for the image of a lensed source to cross a GMC, and thus being periodically magnified via microlensing on individual clumps, scales with the projected area of the GMCs. 
It is a factor $\sim60$ larger then the value obtained in Eq.~\ref{eq6}.

In fact, there are reasons to believe that at least one of the images of \srcs\ crosses a GMC in the lensing galaxy.
\cite{fa99} interpreted the different reddening of the two images of \srcs\ as an additional absorption of the leading image with the differential extinction $\Delta E(B-V)=0.90\pm0.14$. 
The absorption is so strong, that it inverts the brightness ratio of the two images in the optical range, making the trailing image brighter. 
Moreover, molecular absorption line has been detected in the leading image, allowing the estimation of the H$_2$ column density, which is a rather large value of $0.5-5\times 10^{22}\mathrm{cm^{-2}}$ \citep{mr96}. 
Interestingly, a similarly large column of absorbing gas has been also detected in the other known gravitationally lensed gamma-ray quasar PKS1830-211 \citep{wc96}, for which microlensing was suggested \citep{nvm15}.

\section{Magnification maps on intermediate scale structures}\label{sec:magn}
In this section we use numerical simulations to compute the magnification maps caused by microlensing on intermediate scale structures.
We use the inverse ray shooting method (see e.g. \citealp{sw86}).
We project a large number of points homogeneously on the lens plane and compute their corresponding position in the source plane.
To compute the deflection angles of individual rays we use the typical thin sheath approximation of the lens, i.e. we project the 3D distribution of the mass on the XY plane of the lens. 
Then, we divide the source plane in a grid of cells and compute the magnification as the ratio of the number of rays hitting a given cell to the average number of rays emitted in the solid angle of this cell. 
\kom{The maps presented in this work are divided in 1000x1000 cells and  $1-5\times10^8$ rays are simulated per map.}
Therefore, the accuracy of the computed multiplications is $\sim 5-10\%$.
The resolution of the maps is defined as the linear size of the cells in which the magnification is computed.
For larger sources, the magnification would average over multiple cells of the map, resulting in smaller values for sharp features. 

\subsection{Microlensing on a GC}
\begin{figure*}
\includegraphics[width=0.49\textwidth]{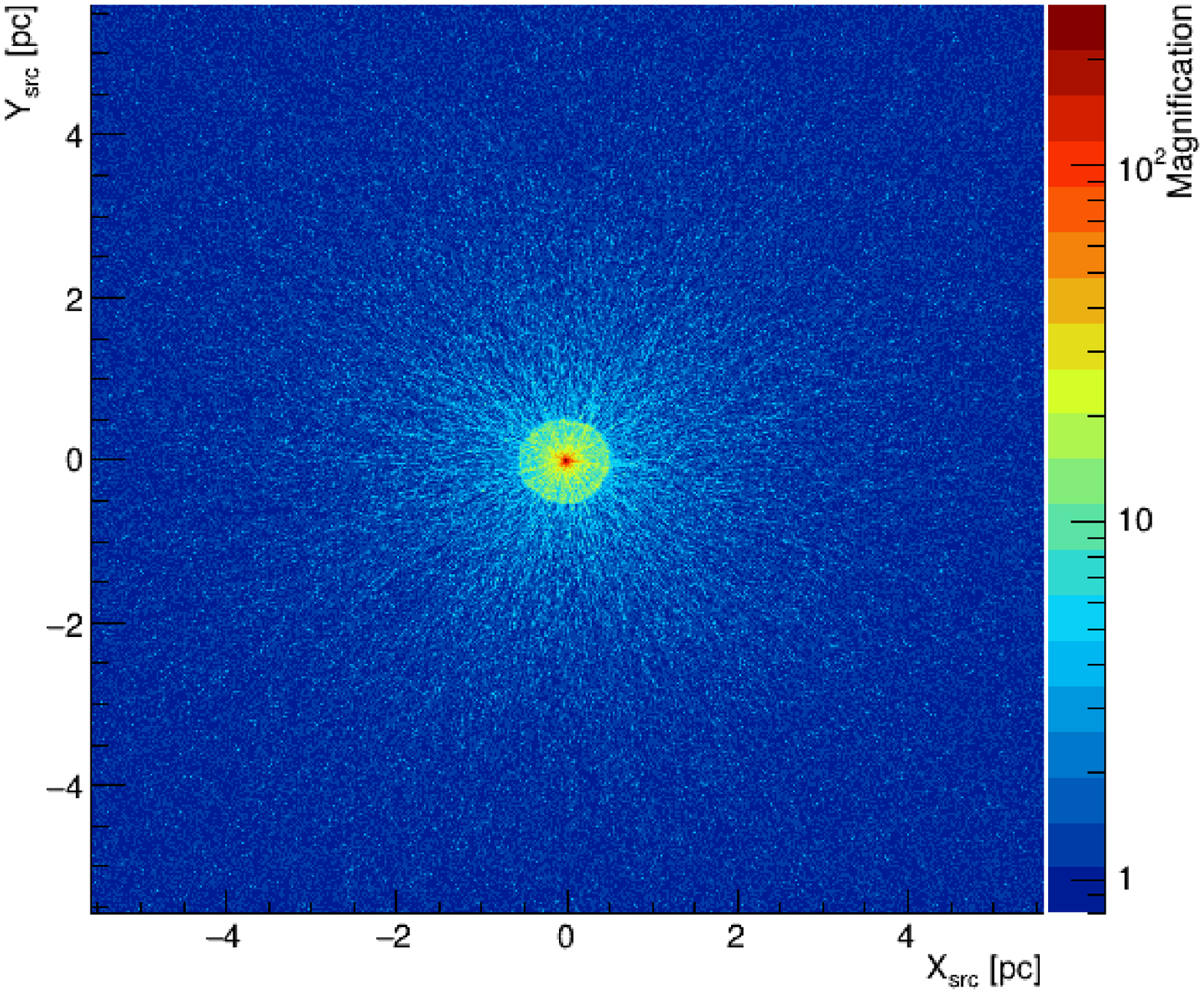}
\includegraphics[width=0.49\textwidth]{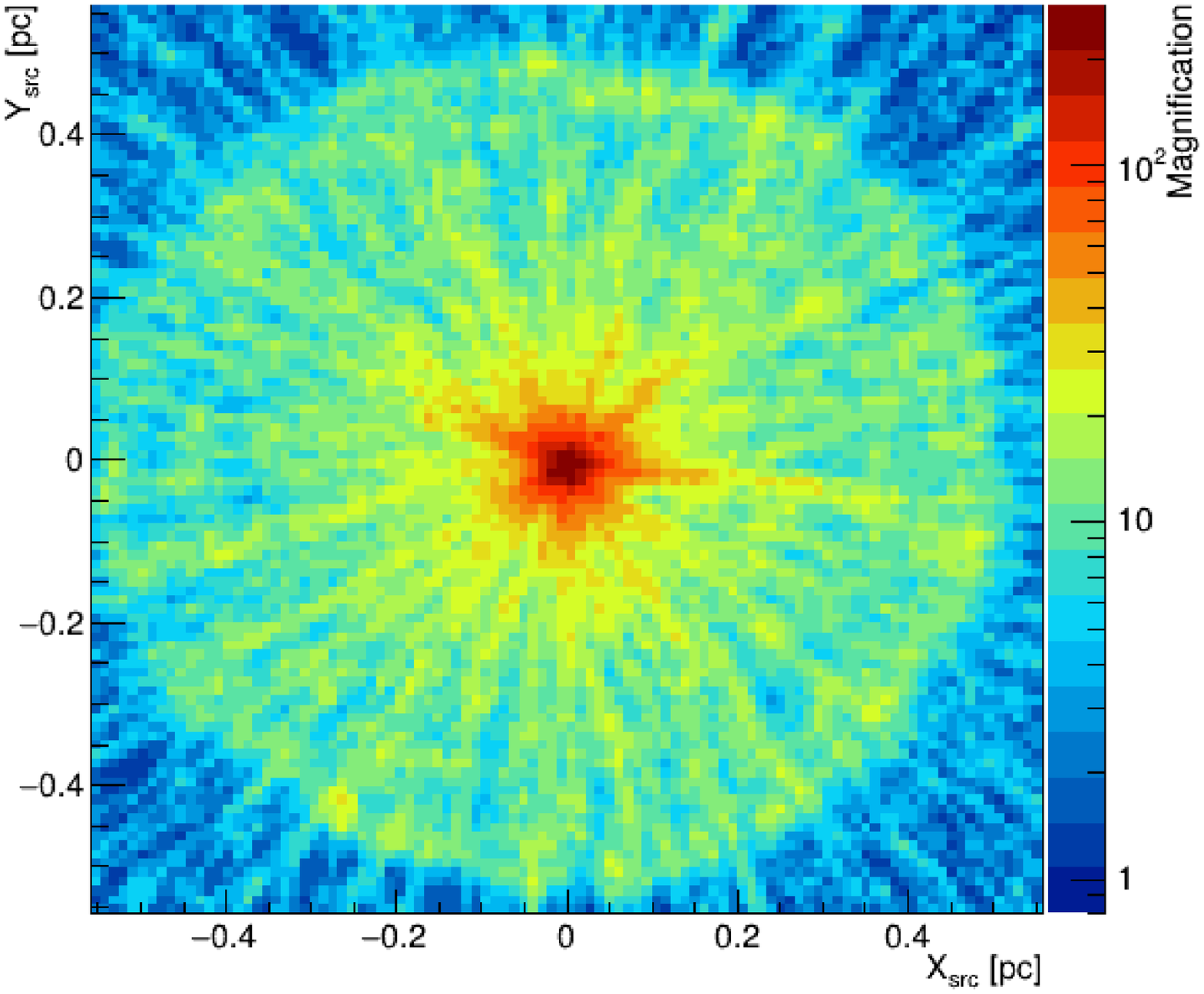} \\
\includegraphics[width=0.43\textwidth]{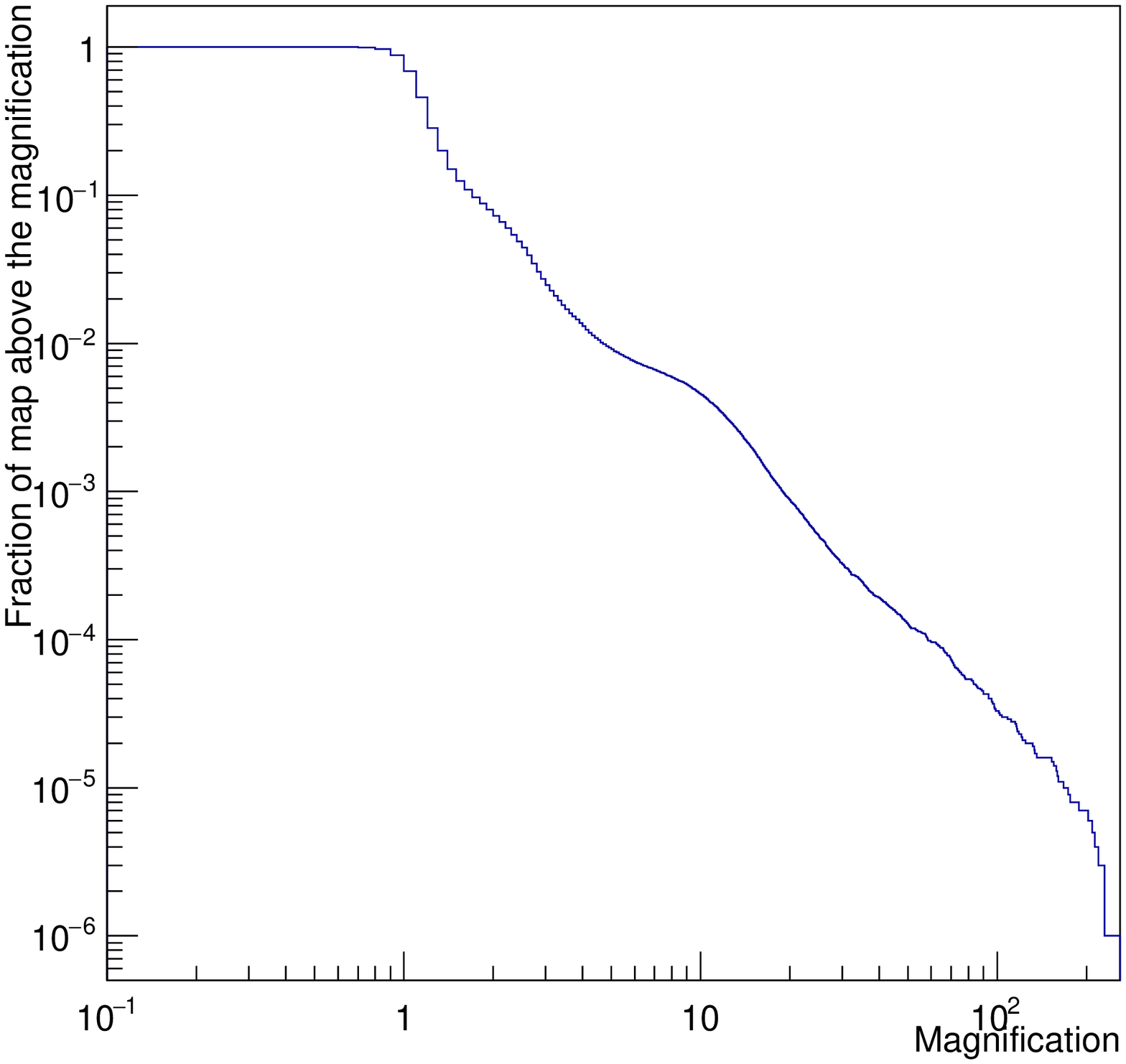}
\includegraphics[width=0.49\textwidth]{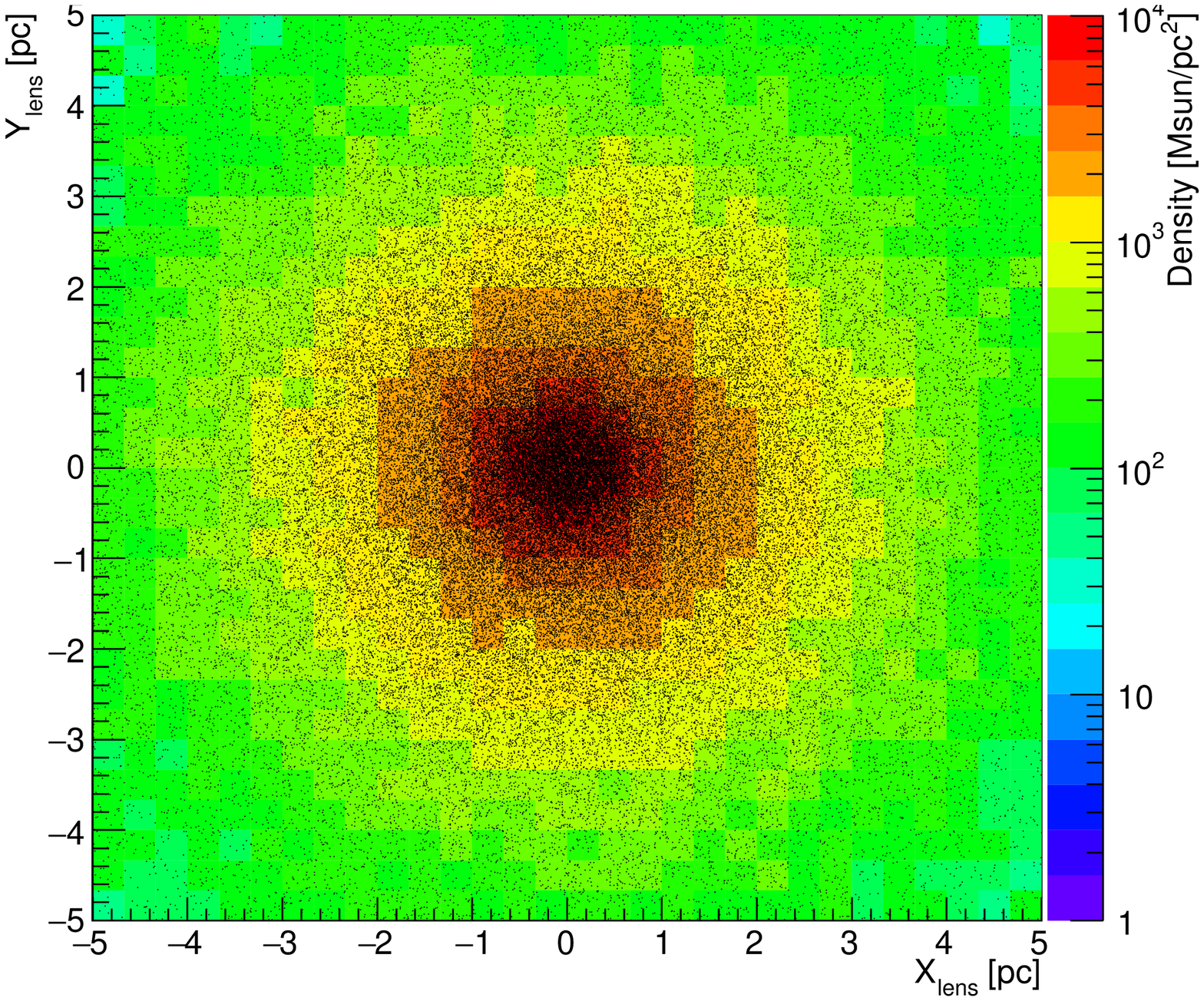}
\caption{Microlensing by a GC with $M_{\rm GC}=10^5 M_\odot$, $R_{\rm c}=0.5\,$pc and $R_{\rm h}=3\,$pc.
Top panels: magnification maps obtained for the resolution of 0.011\,pc (right map is zoomed by a factor of 10)
Fraction of a map with a magnifications above a given value is shown in the bottom left panel.
The surface mass density of GC (in the reference frame of the lens), with individual stars marked with black points is shown in bottom right panel. 
}\label{fig:gc1}
\end{figure*}

We simulate a GC with the total mass of $10^{5} M_\odot$ composed of individual stars with masses of $0.8 M_\odot$.
The stars are distributed according to Eq.~\ref{eqgc}.
Note that as the number of stars is large, and we are interested in the lensing on the whole structure of GC rather than on the individual stars the precise distribution of the star masses is not relevant for this study. 
We select $R_{\rm c}=0.5\,$pc and $R_{\rm t}=30\,$pc (corresponding to $R_{\rm h}=3\,$pc) as a typical GC parameters. 
The shadowing of individual rays by the stars is negligible, therefore the stars are treated as point masses.
In Fig.~\ref{fig:gc1} we show the magnification map obtained from such a GC with the inverse ray shooting method. 
The caustics normally occurring in the microlensing on a set of point-like masses are not visible in this plot as they happen on size scales below the resolution of the plot.
On the other hand, the combined effect of the stars in the core of the GC causes strong magnification.
Values of the order of 10 occur if the projection of the emission region crosses the core.
For a source with an emission region below $0.01$\,pc, magnifications up to a factor of 100 are achieved at the centre of the GC. 

\subsection{Microlensing on an OC}
OCs are less abundant in stars than GCs.
They also show higher irregularities in their structure.
This causes difficulties in modelling the distribution of the mass inside OC. 
For those simplified calculations we assume the same profile as was used in the case of GCs (see Eq.~\ref{eqgc}). 
We selected parameters describing the typical size of an OC following the \cite{kh05} catalog. 
We simulate the OC with a geometric size determined by $R_c=1.5$\,pc and $R_t=4.5$\,pc and a total mass of $M_{\rm OC}=3\times10^3 M_\odot$. 
We assumed a power-law distribution of the mass of the stars with a standard index of $-2.3$.
The distribution spreads between $0.1\,M_\odot$ and $100\,M_\odot$.
Those mass ranges were selected such that the number of massive OB stars agrees (after correction for the total mass of the OC) with the number observed in the Cyg OB2 cluster \citep{bu03}.

\begin{figure*}
\includegraphics[width=0.34\textwidth]{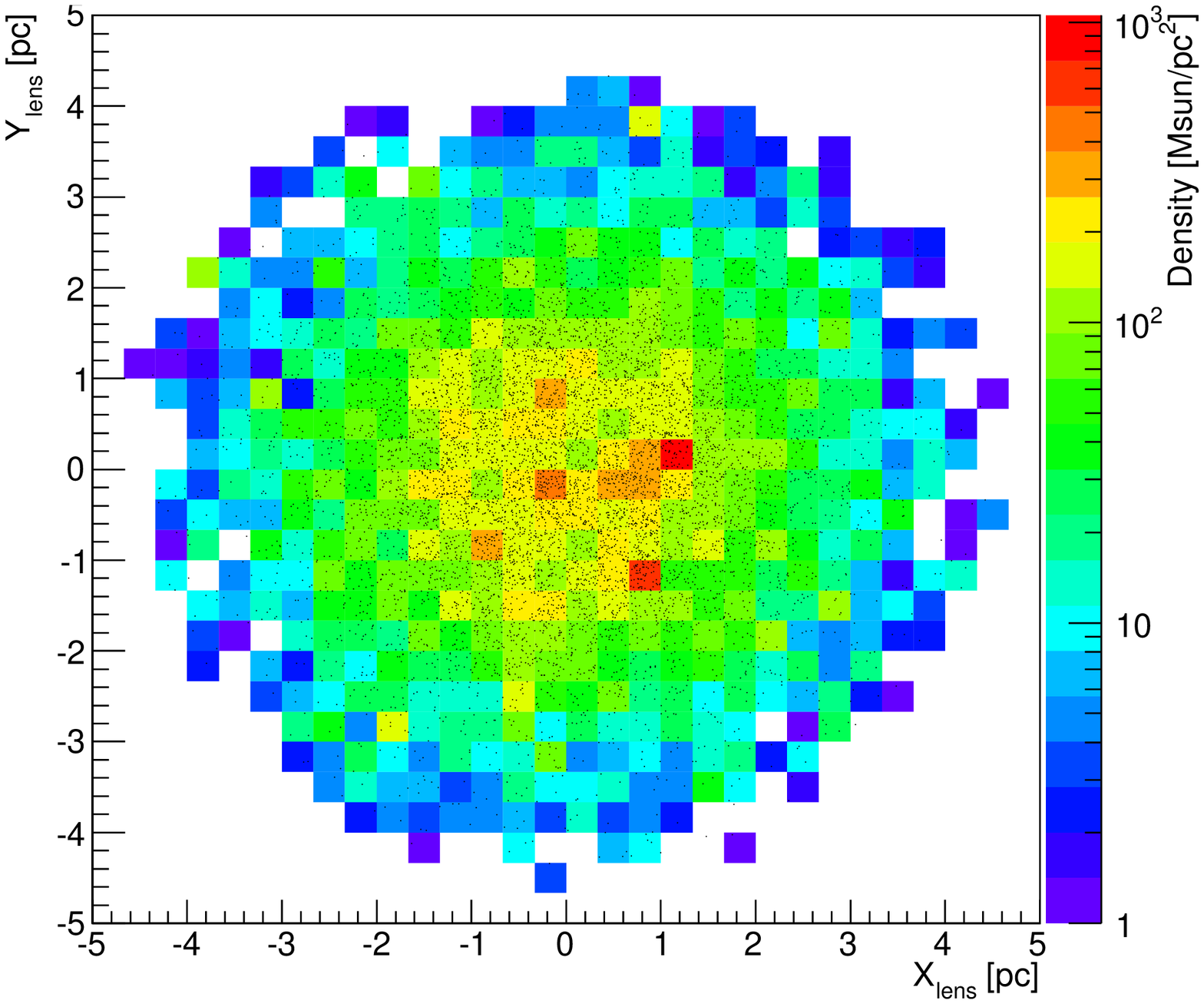}
\includegraphics[width=0.34\textwidth]{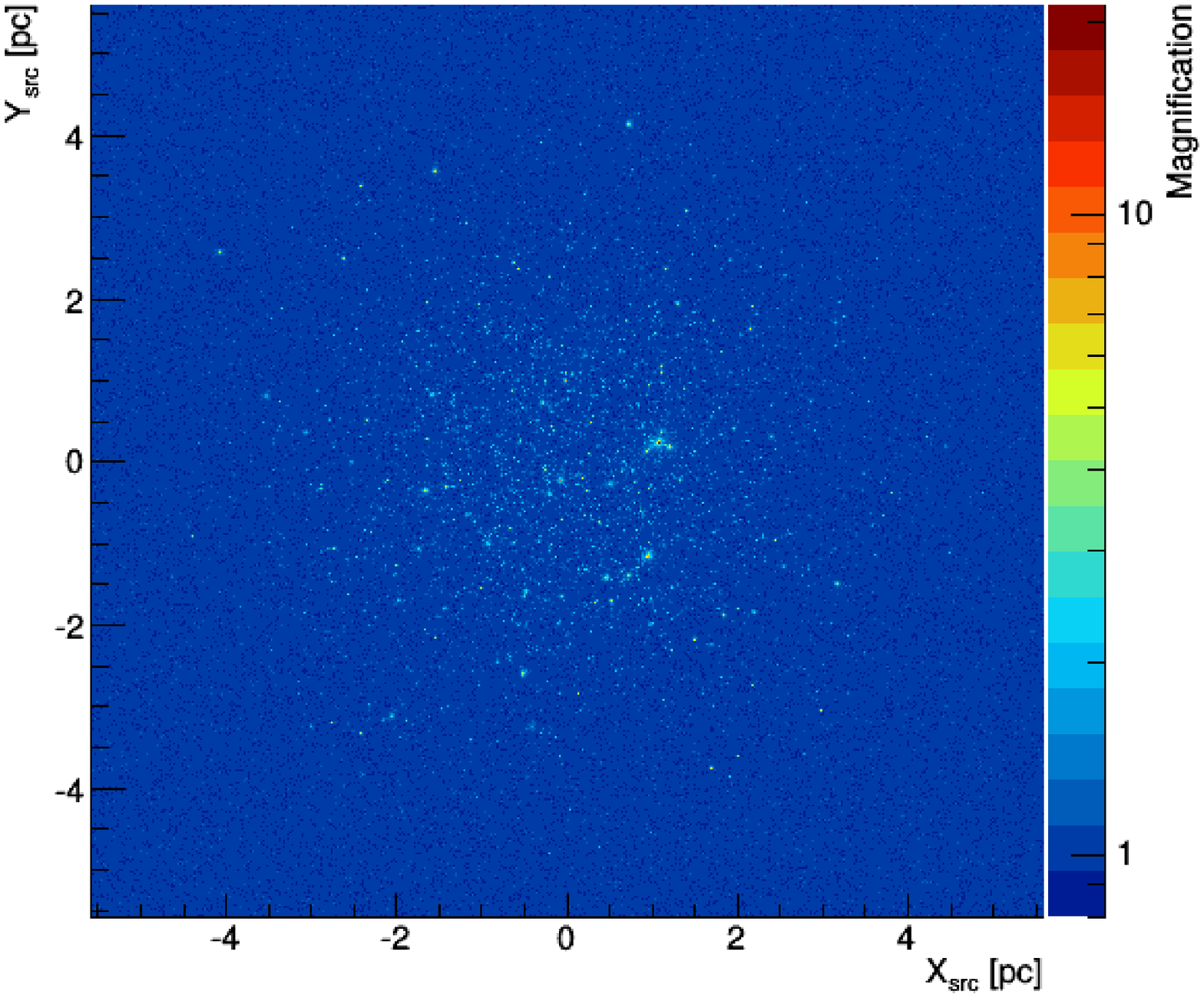}
\includegraphics[width=0.30\textwidth]{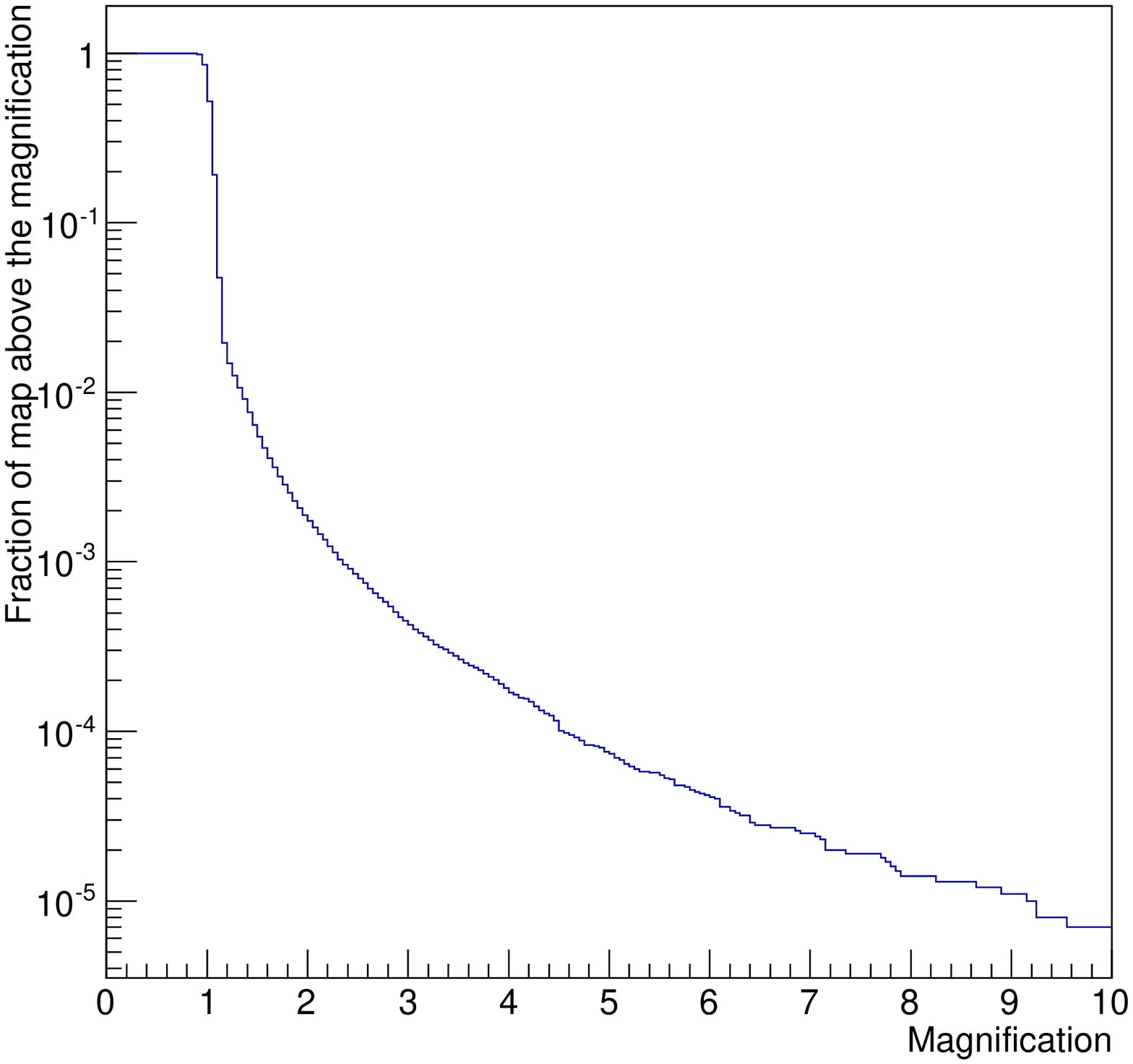}
\caption{Microlensing by an OC with $M_{\rm OC}=3\times10^3 M_\odot$, $R_{\rm c}=0.5\,$pc and $R_{\rm t}=4.5\,$pc.
The surface mass density of the OC (in the reference frame of the lens), with individual stars marked with black points (the left panel).
Magnification map obtained for the resolution of 0.011\,pc are shown in the middle panel.
Fraction of the map with a magnification larger than a given value is shown in the right panel.
}\label{fig:oc1}
\end{figure*}

The magnification obtained in such a case is presented in Fig.~\ref{fig:oc1}.
As OCs are on average much less massive than GCs and still spread over a relatively large size, there is nearly no coherent lensing by the whole structure. 
Relatively small regions, 0.01\,pc (resolution of the simulations) -- 0.05\,pc with a magnification a factor of a few occur due to the microlensing on individual stars and crossing of caustics for the case of a few nearby stars.

\subsection{Microlensing on a GMC}
GMCs can have a very complicated structure. 
For simplicity we simulate a GMC as a spherical structure composed of individual extended clumps. 
We use a typical mass and radius of GMC of $M_{\rm GMC}=2\times10^5 M_\odot$, $R_{\rm GMC}=20\,$pc. 
We consider two different scenarios of the distribution of clumps, homogeneous (i.e. $dN_{\rm clump}/dV\propto \mathrm{const}$) and more peaked towards the centre of the GMC (i.e. $dN_{\rm clump}/dV\propto r^{-1}$). 
We take the mass distribution function of the clumps from \cite{sg90}. 
The masses of individual clumps are drawn from a power law distribution with an index $-1.7$. 
The values of masses are spread between $0.8 M_\odot$ and $3\times10^3 M_\odot$.
The radius of the clump is estimated following the empirical correlation shown by \cite{sg90}:
\begin{equation}
R_{\rm clump} = 0.22\mathrm{pc} \times \sqrt{M_{\rm clump}/100 M_\odot}.
\end{equation}
We assume that the individual clumps have homogeneous mass density. 

\begin{figure*}
\centering
\includegraphics[width=0.34\textwidth]{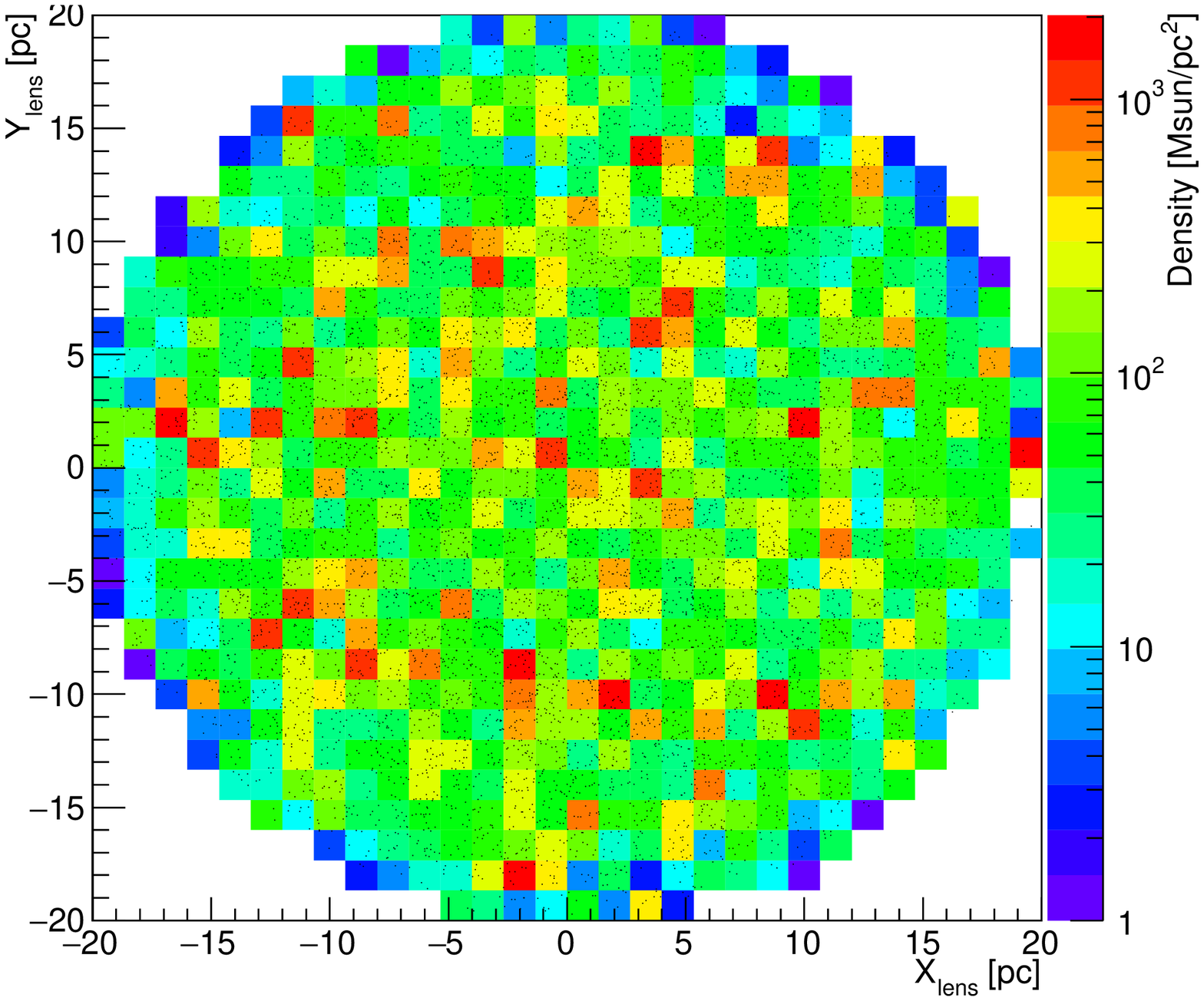}
\includegraphics[width=0.34\textwidth]{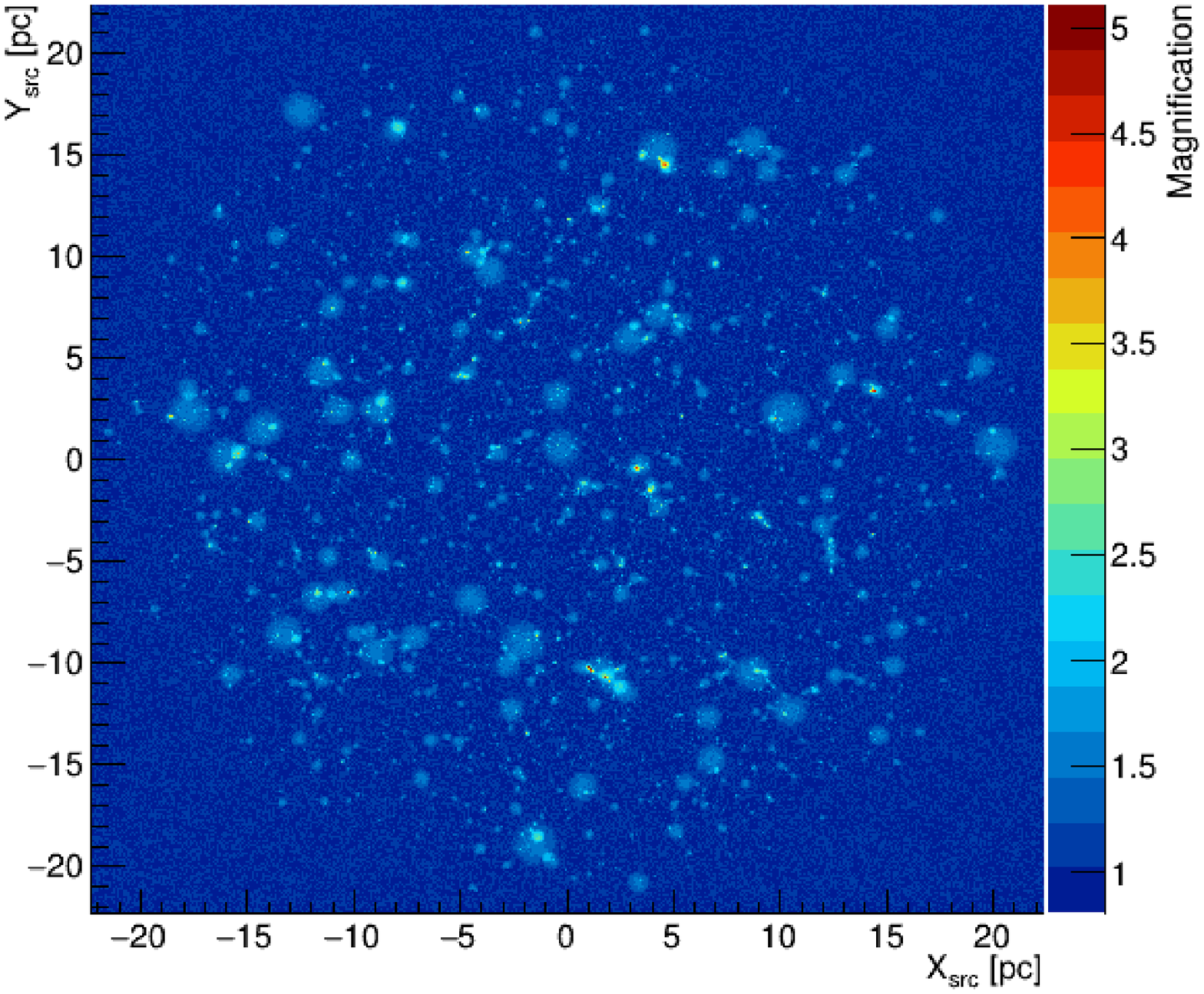}
\includegraphics[width=0.30\textwidth]{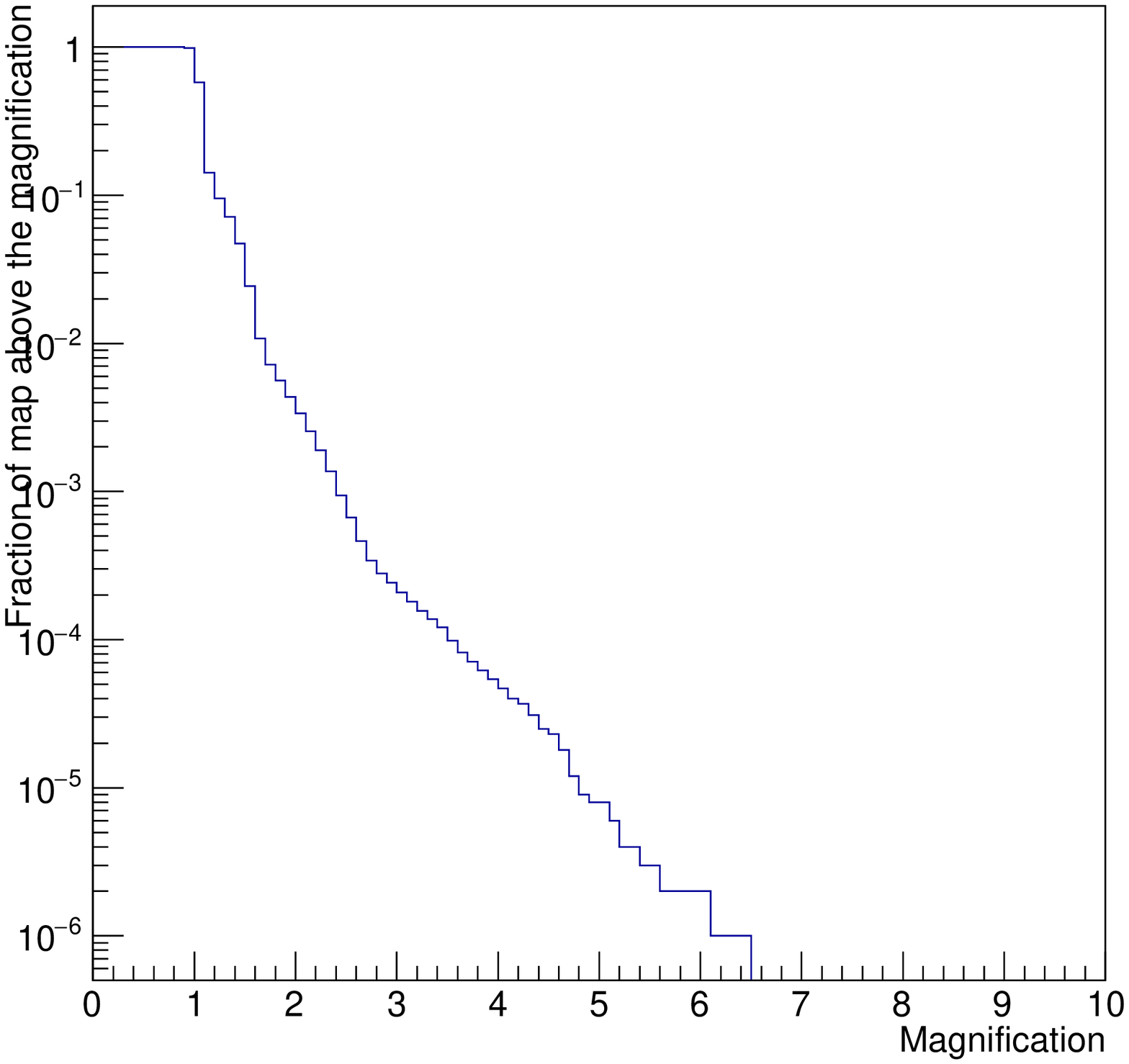} \\
\includegraphics[width=0.34\textwidth]{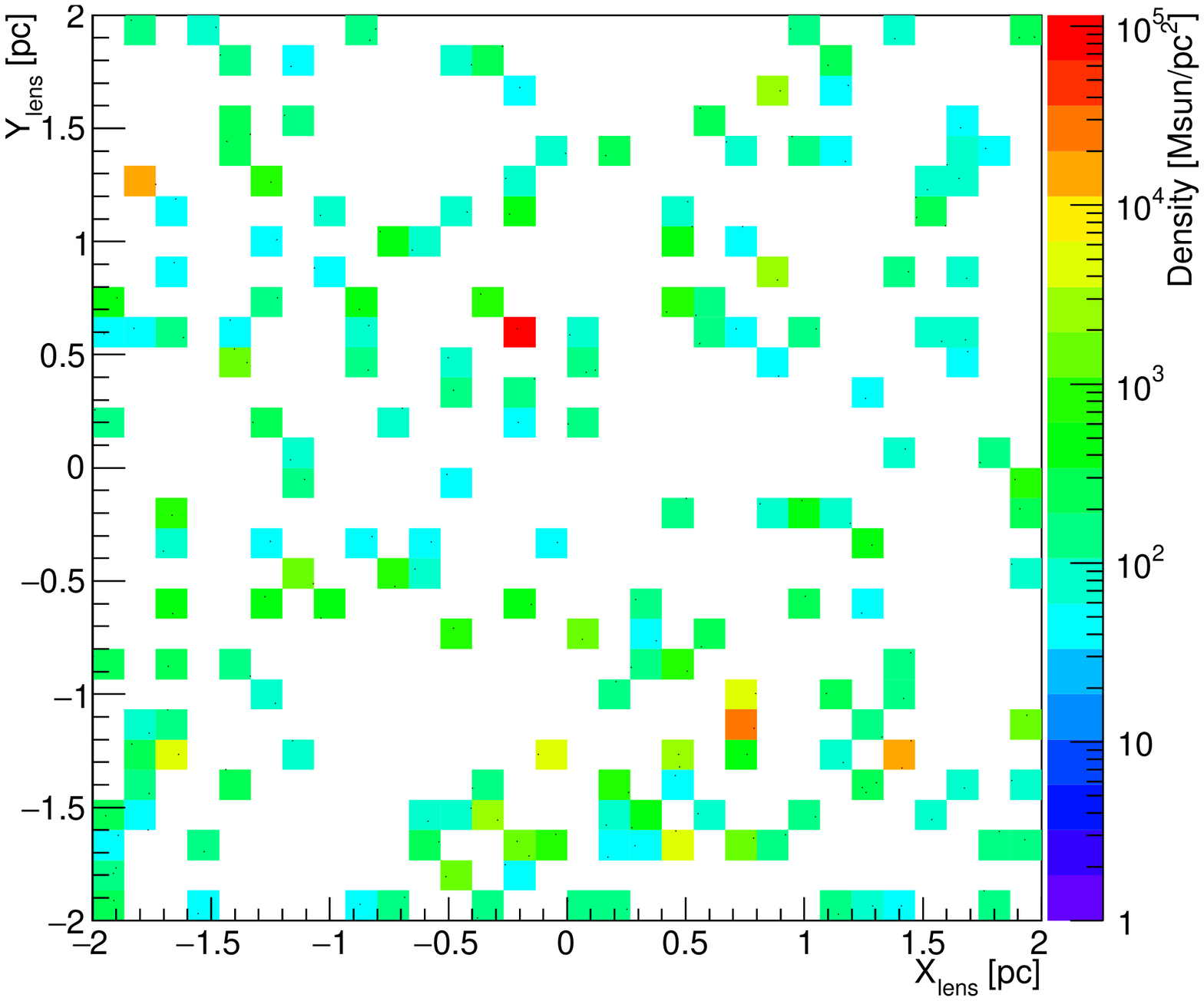}
\includegraphics[width=0.34\textwidth]{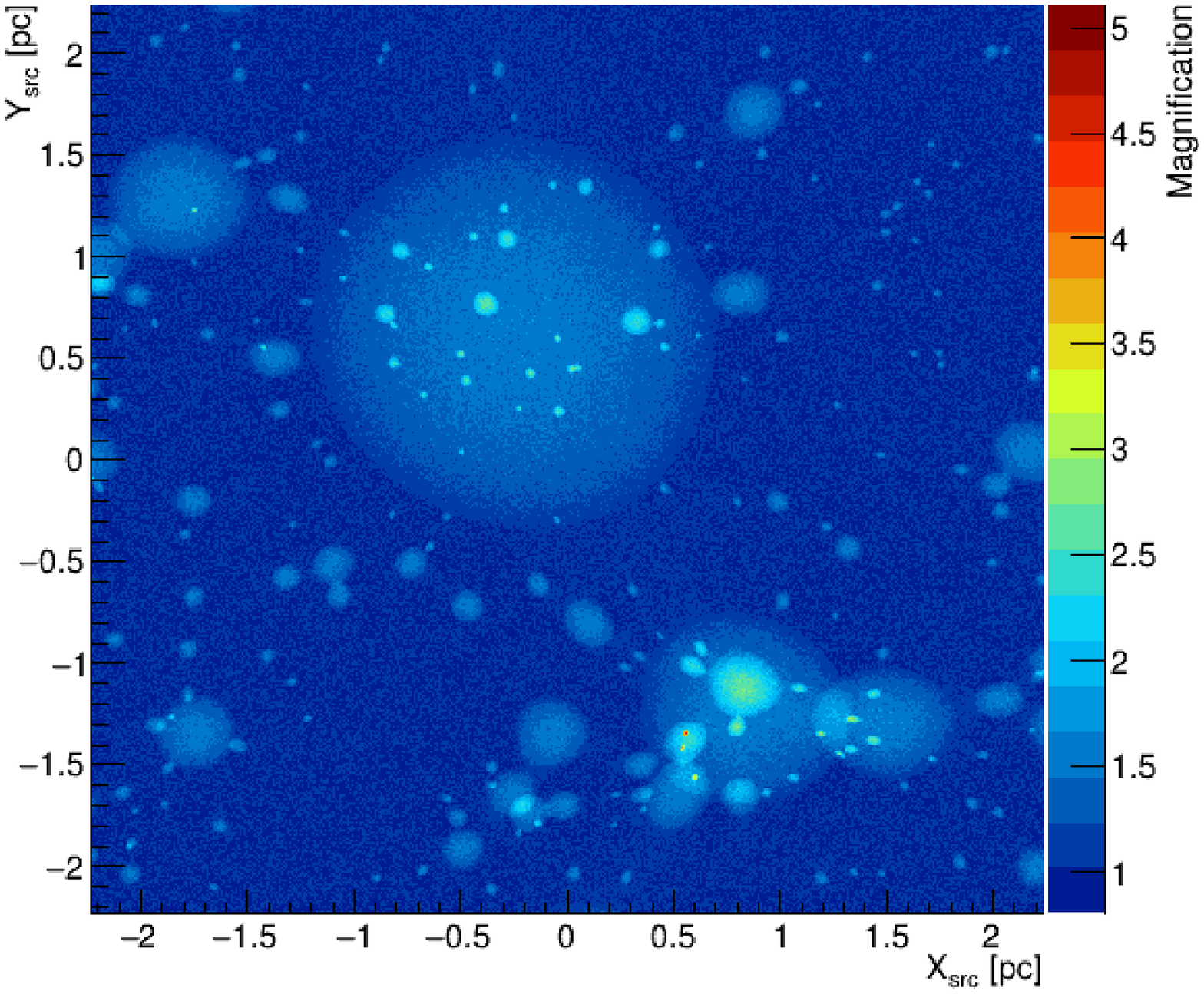}
\includegraphics[width=0.30\textwidth]{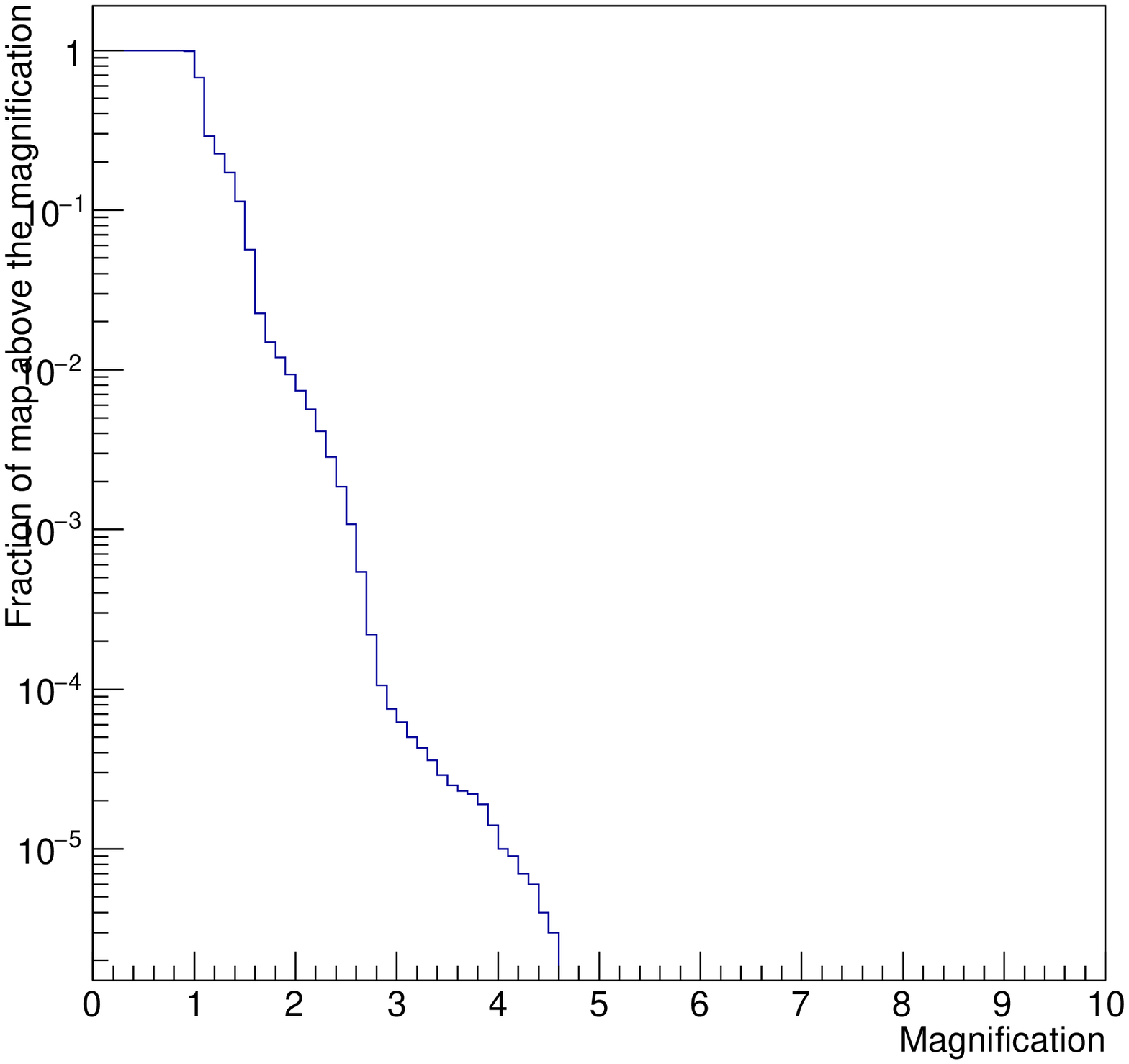}
\caption{Microlensing by a GMC with $M_{\rm GMC}=2\times10^5 M_\odot$, $R_{\rm GMC}=20\,$pc composed of homogeneously distributed clumps.
Surface density (in the reference frame of the lens) of the mass of the lens, with individual clumps marked with black points (left panels). 
Magnification map in the reference frame of the source (middle panels).  
Fraction of the map with a magnification above a given value (right panels)
Top panels show the case of the whole GMC (map resolution of 0.045\,pc)
Zoom to the inner part of the GMC with a higher resolution of 0.0045\,pc is shown in bottom panels.
}\label{fig_gmc1}
\end{figure*}

In Fig.~\ref{fig_gmc1} we show the results of the calculations for the case of homogeneously distributed clumps. 
For the assumed parameters, the Einstein radius computed according to Eq.~\ref{eq:re} is a factor of a few smaller than the actual size of a GMC.
Then, most of the magnification happens on individual clumps. 
Magnifications by a factor of a few can be achieved and occur on distance scales of a fraction of pc.   

\begin{figure*}
\includegraphics[width=0.34\textwidth]{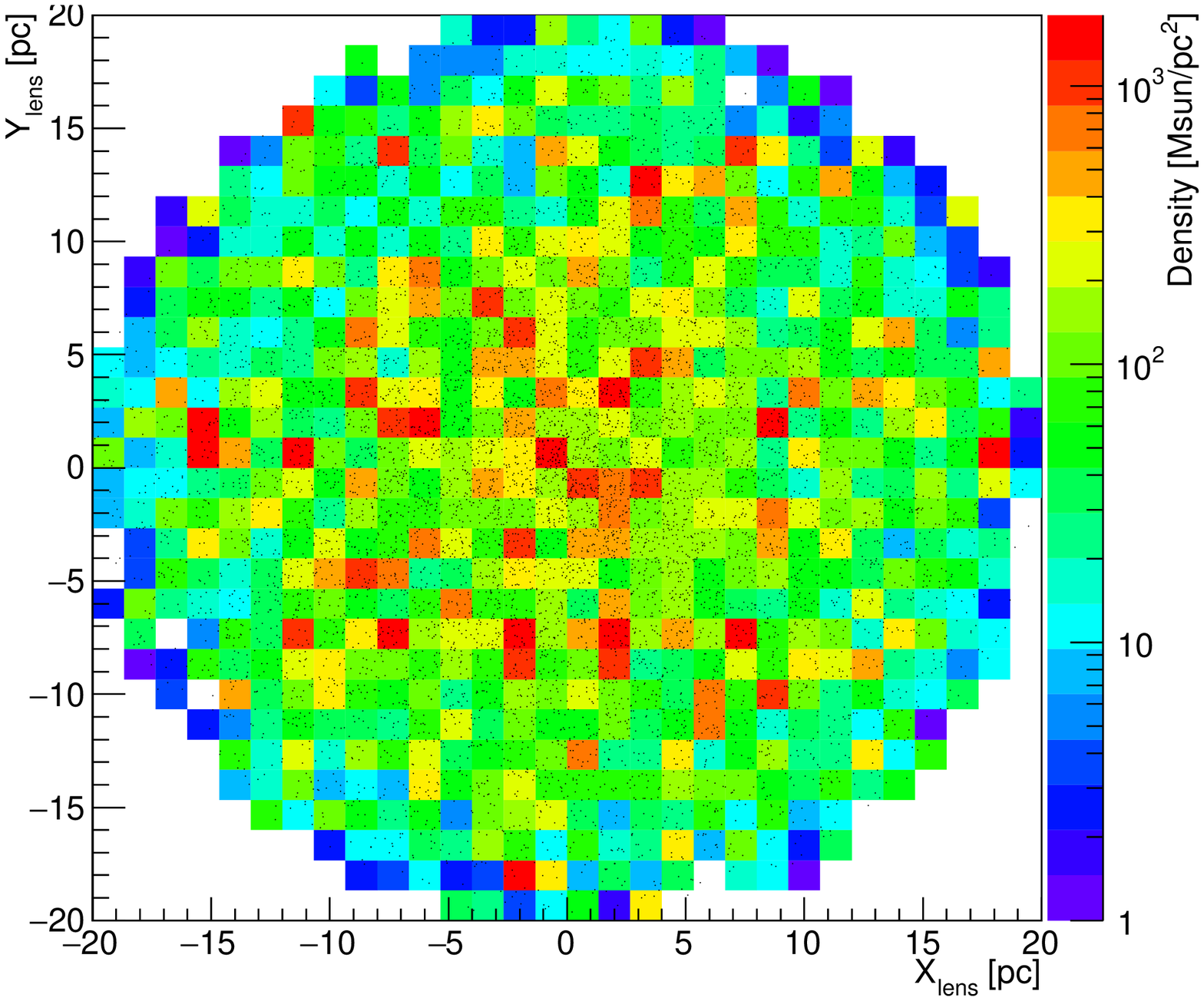}
\includegraphics[width=0.34\textwidth]{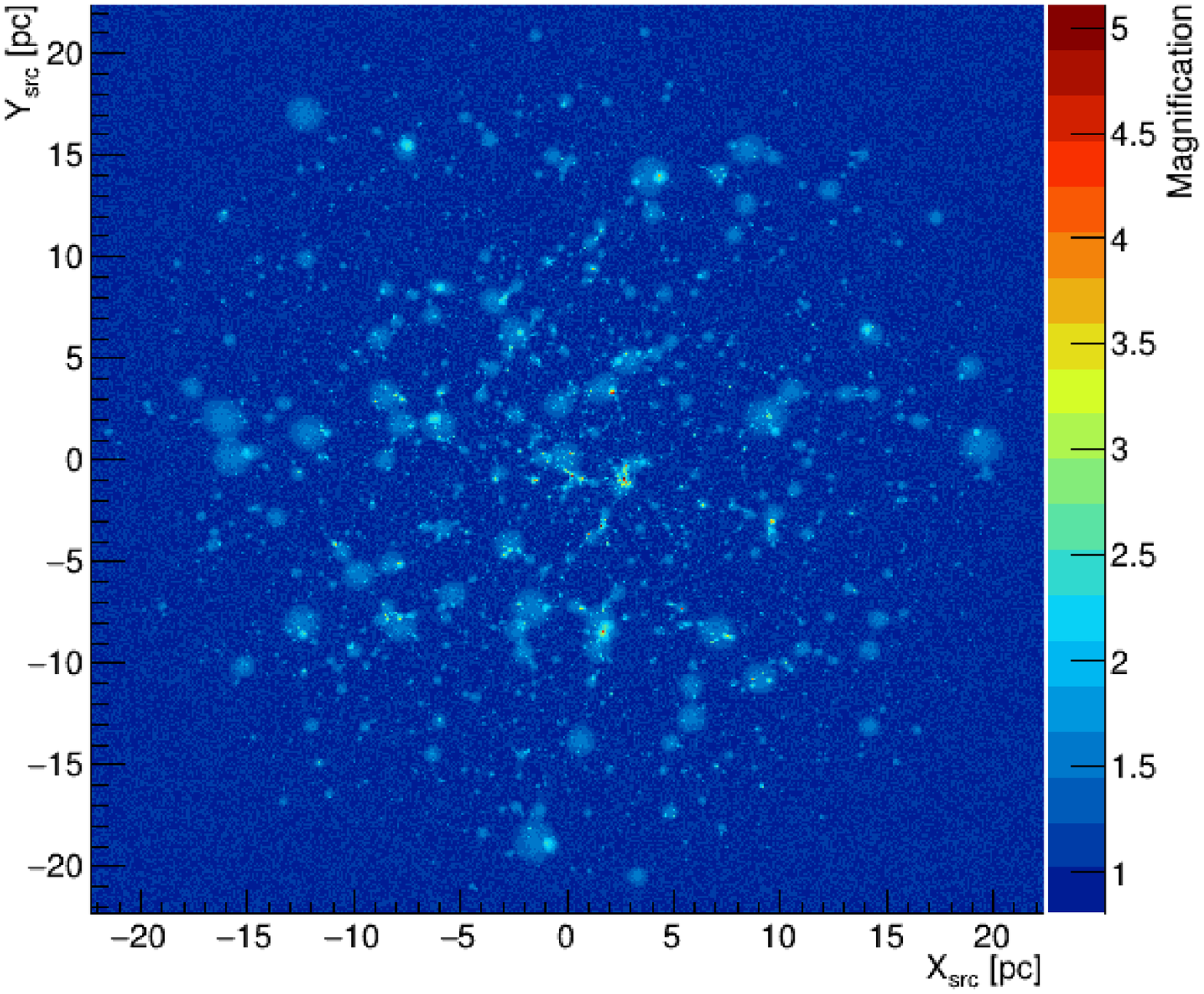}
\includegraphics[width=0.30\textwidth]{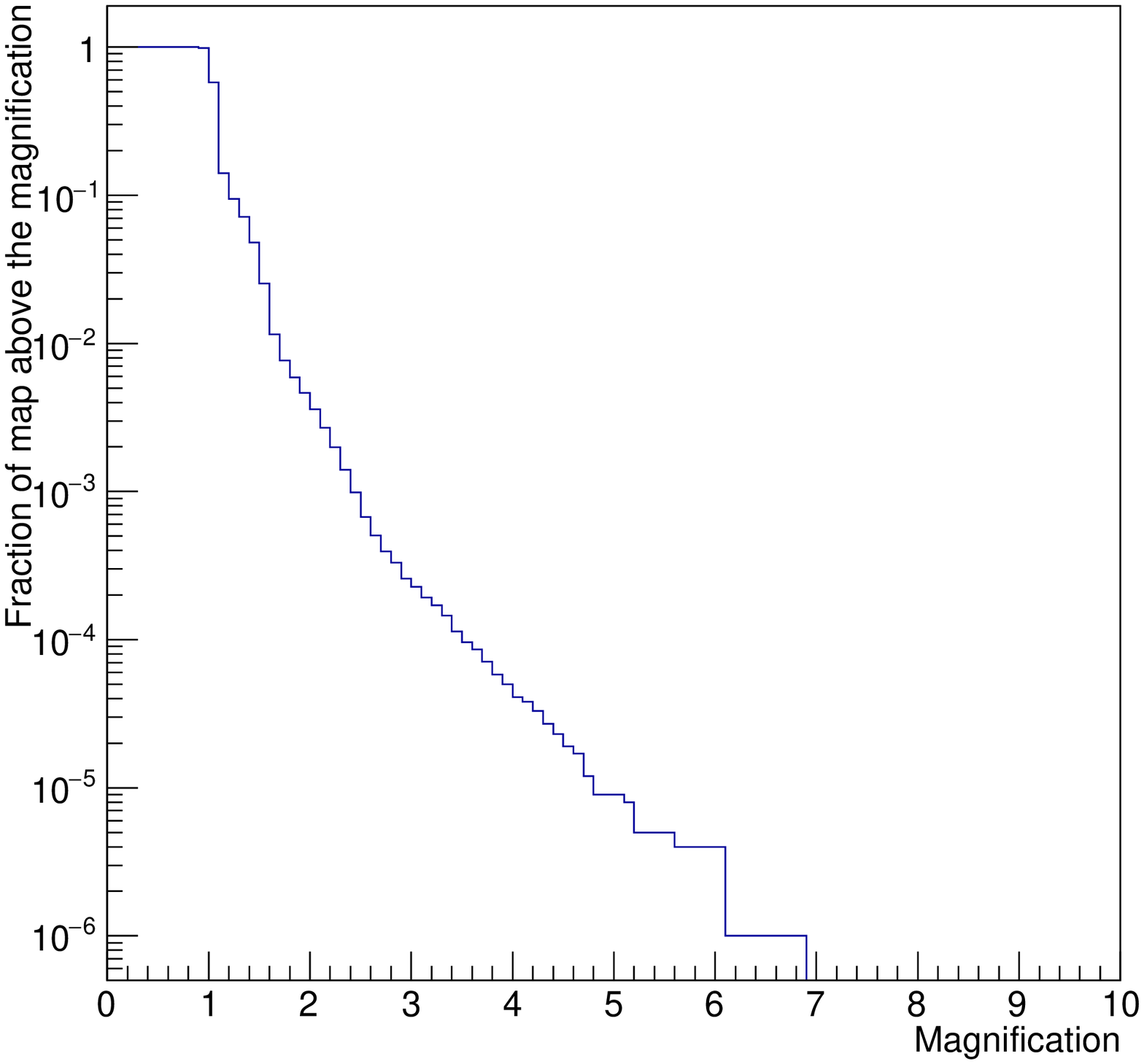}\\
\includegraphics[width=0.34\textwidth]{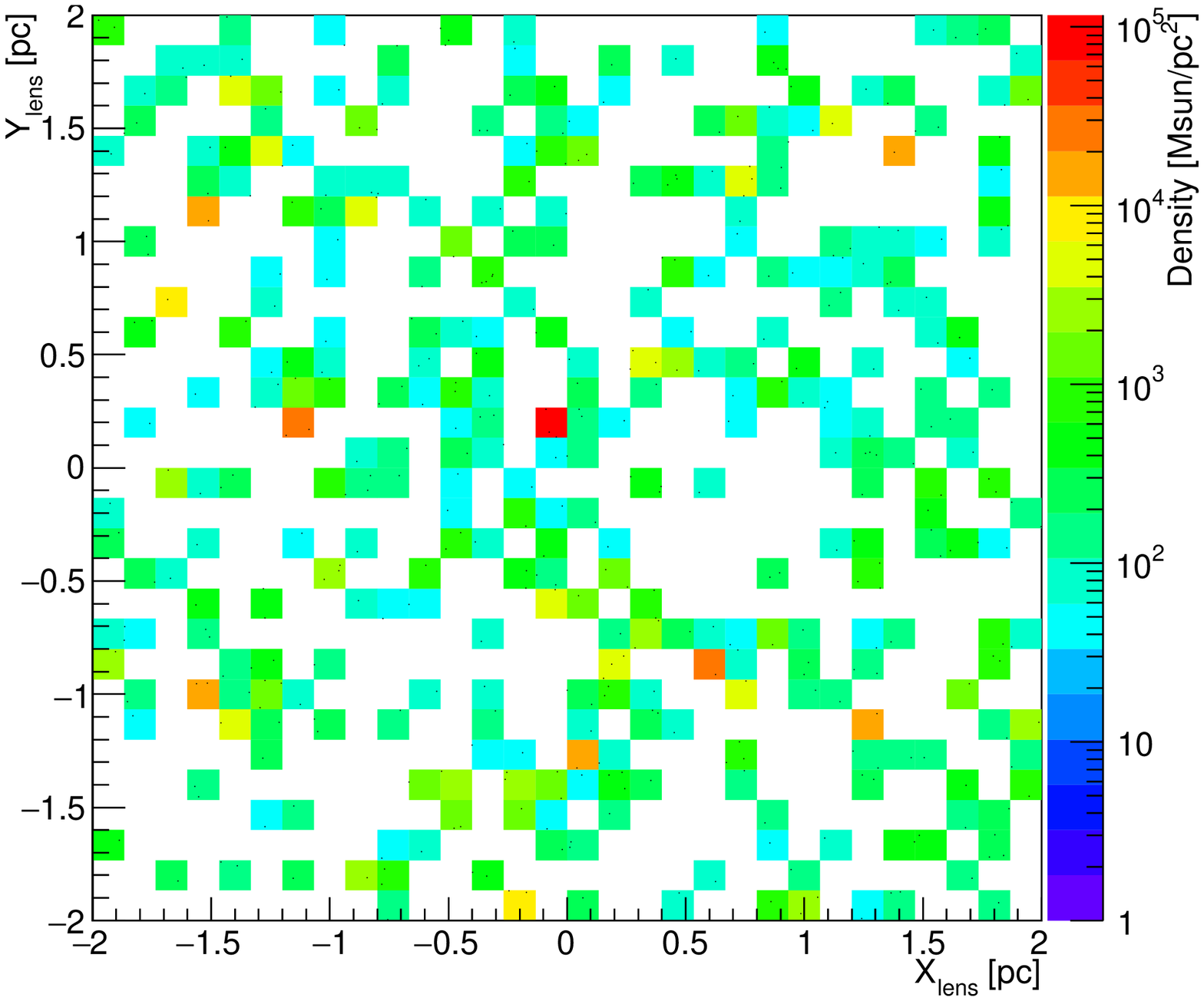}
\includegraphics[width=0.34\textwidth]{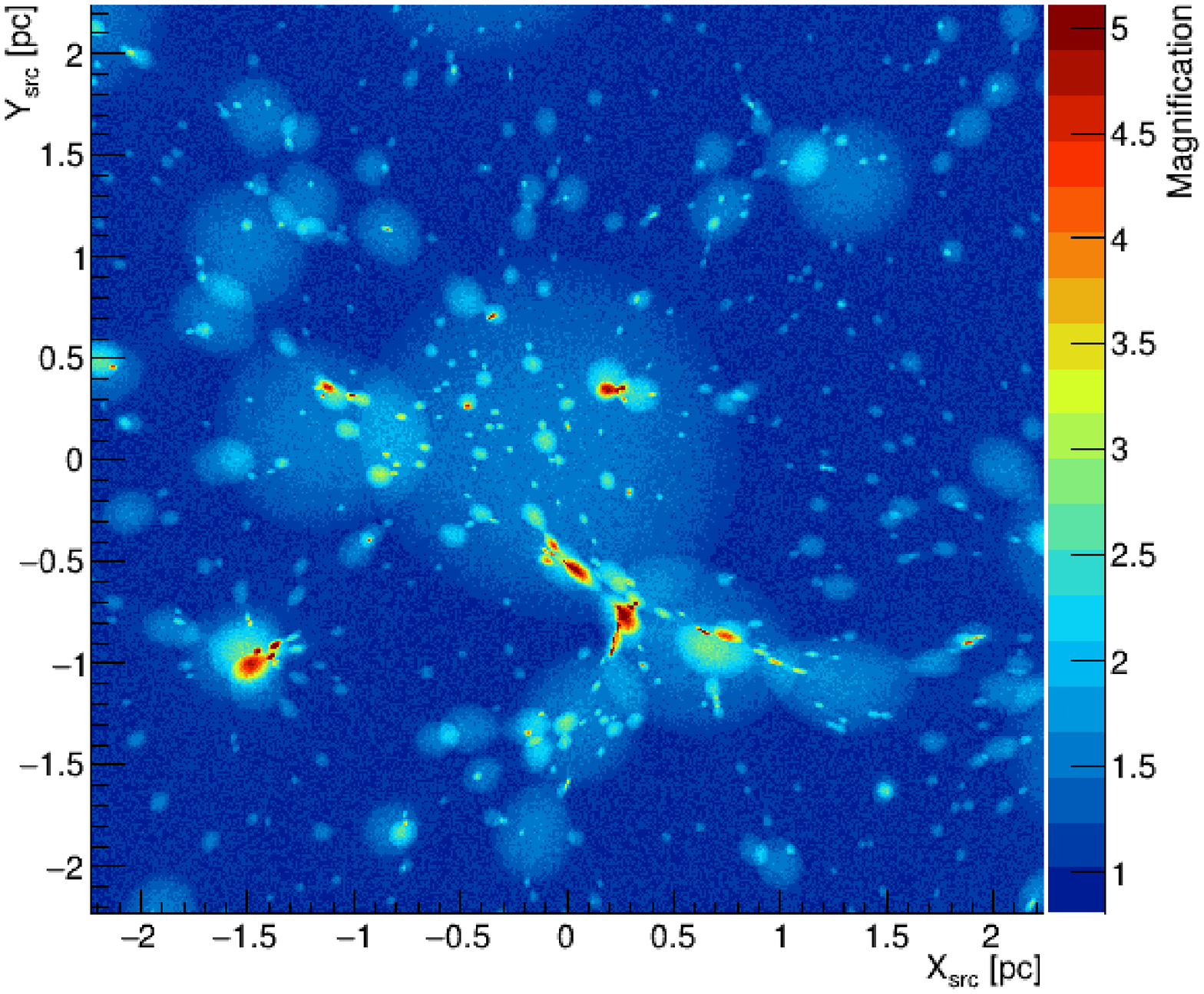}
\includegraphics[width=0.30\textwidth]{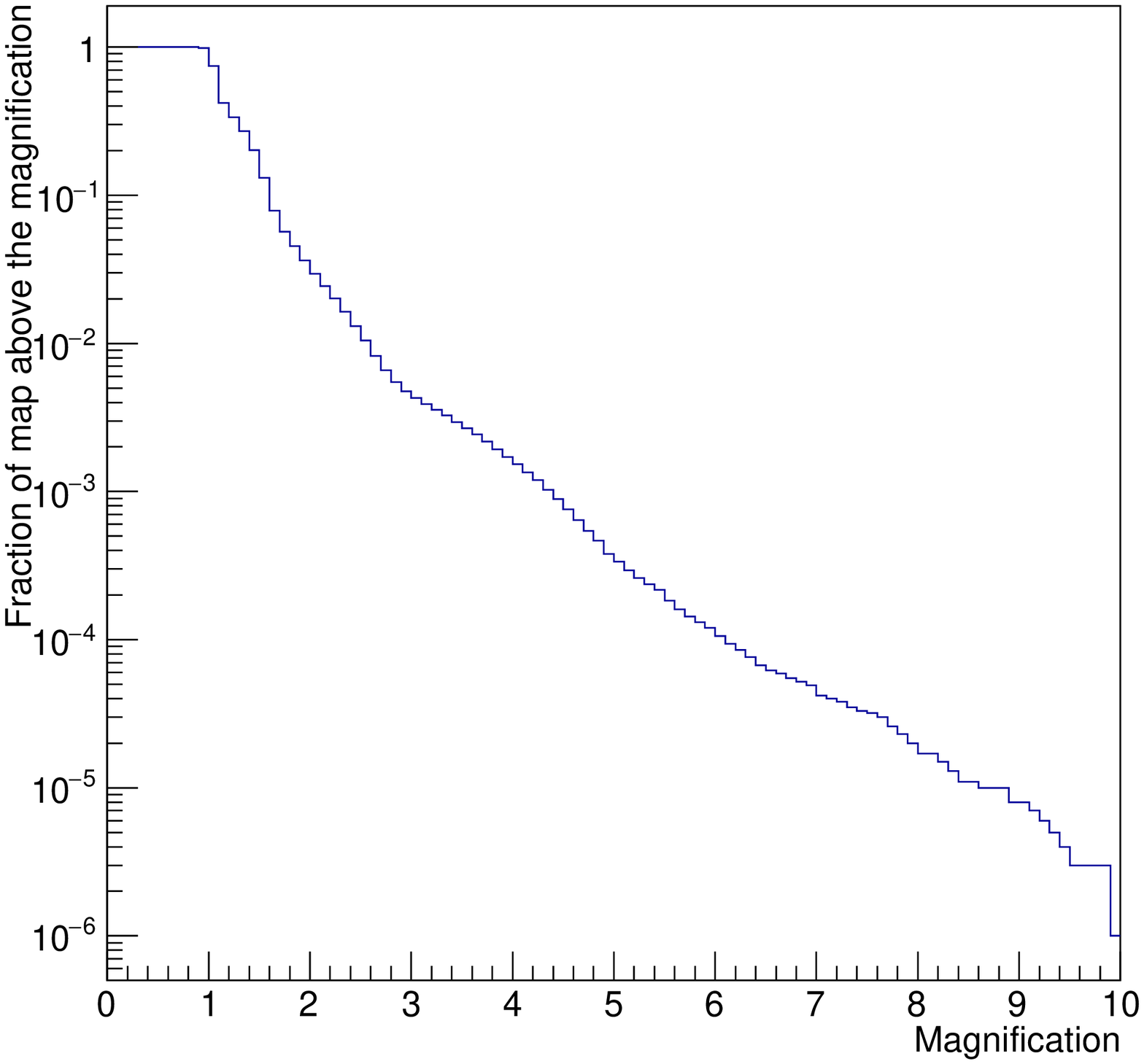}
\caption{As in Fig.~\ref{fig_gmc1} but for a distribution of clumps in GMC following $dN_{\rm clump}/dV\propto r^{-1}$}\label{fig_gmc2}
\end{figure*}

In Fig.~\ref{fig_gmc2} we present the magnification map for the case of GMC with a $dN_{\rm clump}/dV\propto r^{-1}$ density of the clumps. 
In this case, due to more peaked mass distribution, the microlensing is mostly pronounced in the inner pc.
Also, it is much more common for a combination of multiple individual clumps to cause strong magnifications ($>5$) at the caustic crossings.
Note however, that those strong magnifications are possible for sources with the sizes up to $\sim 0.01$\,pc.

\section{Discussion}\label{sec:conc}
If the gamma-ray emission of \srcs\ is emitted according to the classical blob-in-jet models, it would be likely accompanied by a superluminal motion of the emission region.
Such movement is at odds with the interpretation of the change of the magnification ratio of the two images in terms of microlensing on individual stars in the galaxy \citep{vn15}. 
The change in the magnification factor can also be explained as microlensing on intermediate scale structures. 
\citet{vn15} claimed changes in the magnification factors occurring in the \textit{Fermi}-LAT data on the time scales of 20 days.
In order to account for these, with a typical $5\,c$ superluminal motion, one requires features in the magnification map with a length of $\sim0.1$\,pc. 
Such features occur naturally due to microlensing in individual clumps of GMCs. 
Note that in this scenario the constraints on the size of the gamma-ray emission region in \srcs\ are much less restrictive (again $\sim0.1$pc) than for the scenario presented in \citet{vn15}. 

\begin{figure}
\includegraphics[width=0.49\textwidth]{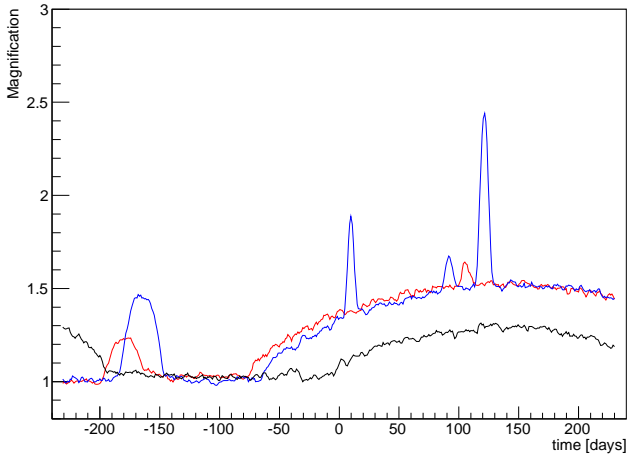}
\caption{
Expected evolution of the magnification shown in the map on Fig.~\ref{fig_gmc1} for a source crossing with superluminal speed of $5\,c$. 
The magnification is averaged over the size of the emission region of 0.01\,pc.
The blue, red and black curves show the case of crossing along the Y direction -0.5\,pc, 0\,pc and +0.5\,pc from the centre of the map
}\label{fig_gmc3}
\end{figure}

\begin{figure}
\includegraphics[width=0.49\textwidth]{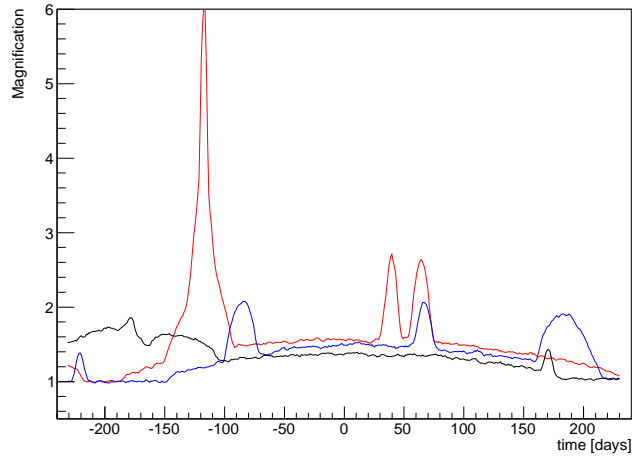}
\caption{
Like in Fig.~\ref{fig_gmc3}, but for the case of GMC with a peaked profile (corresponding to the map in Fig.~\ref{fig_gmc2}).
}\label{fig_gmc4}
\end{figure}

In Fig.~\ref{fig_gmc3} and \ref{fig_gmc4} we show how the magnification evolves with time as the projected position of the emission region crosses through the GMC. 
For those calculations we assume an observed superluminal speed of the source of $5\,c$. 
As expected, crossing of the individual clumps in GMC causes magnification by a factor of a few lasting between days and hundreds of days.
The magnification pattern is complicated as multiple clumps might influence the magnification at a given moment.

\begin{figure}
\includegraphics[width=0.49\textwidth]{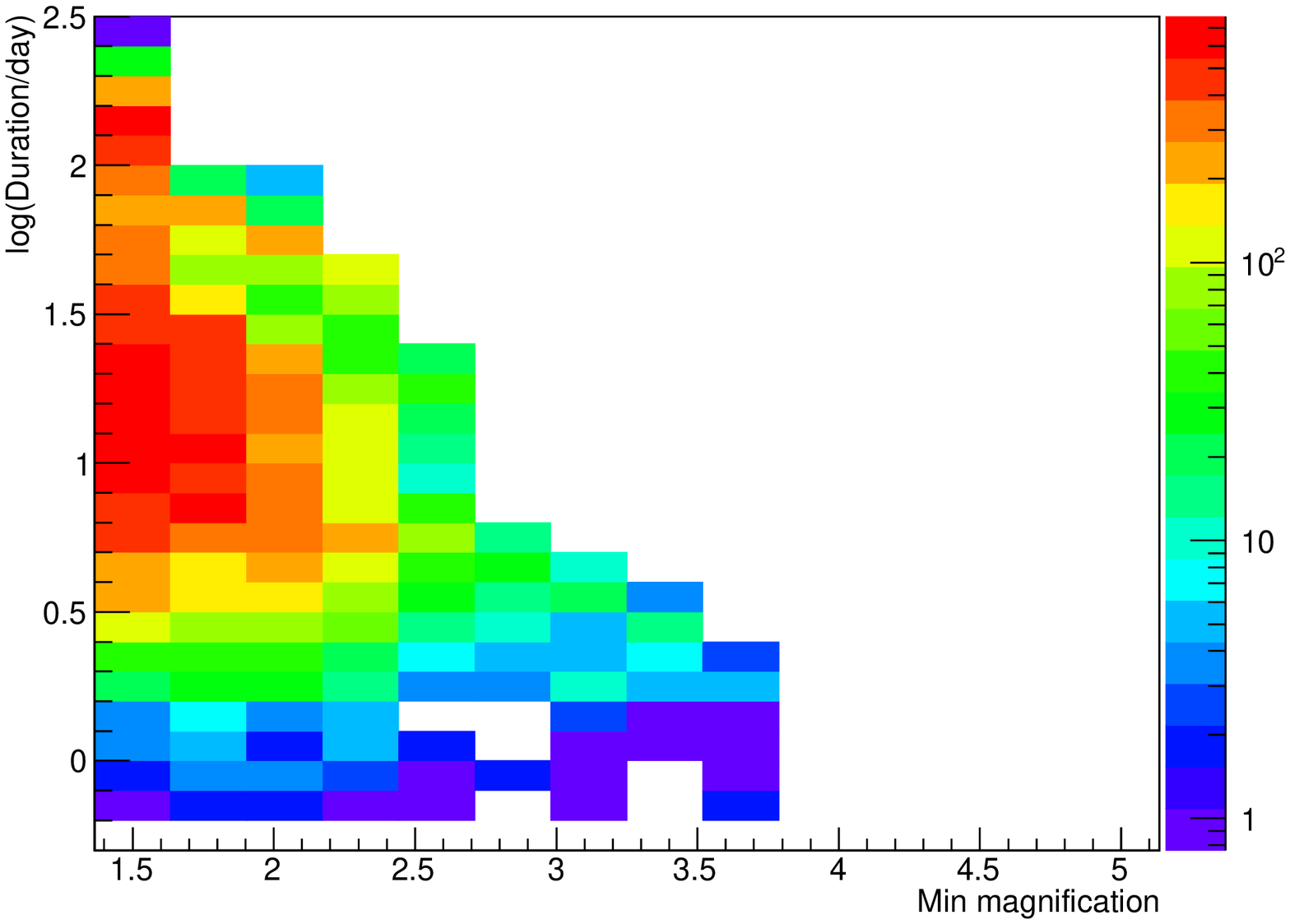}\\
\includegraphics[width=0.49\textwidth]{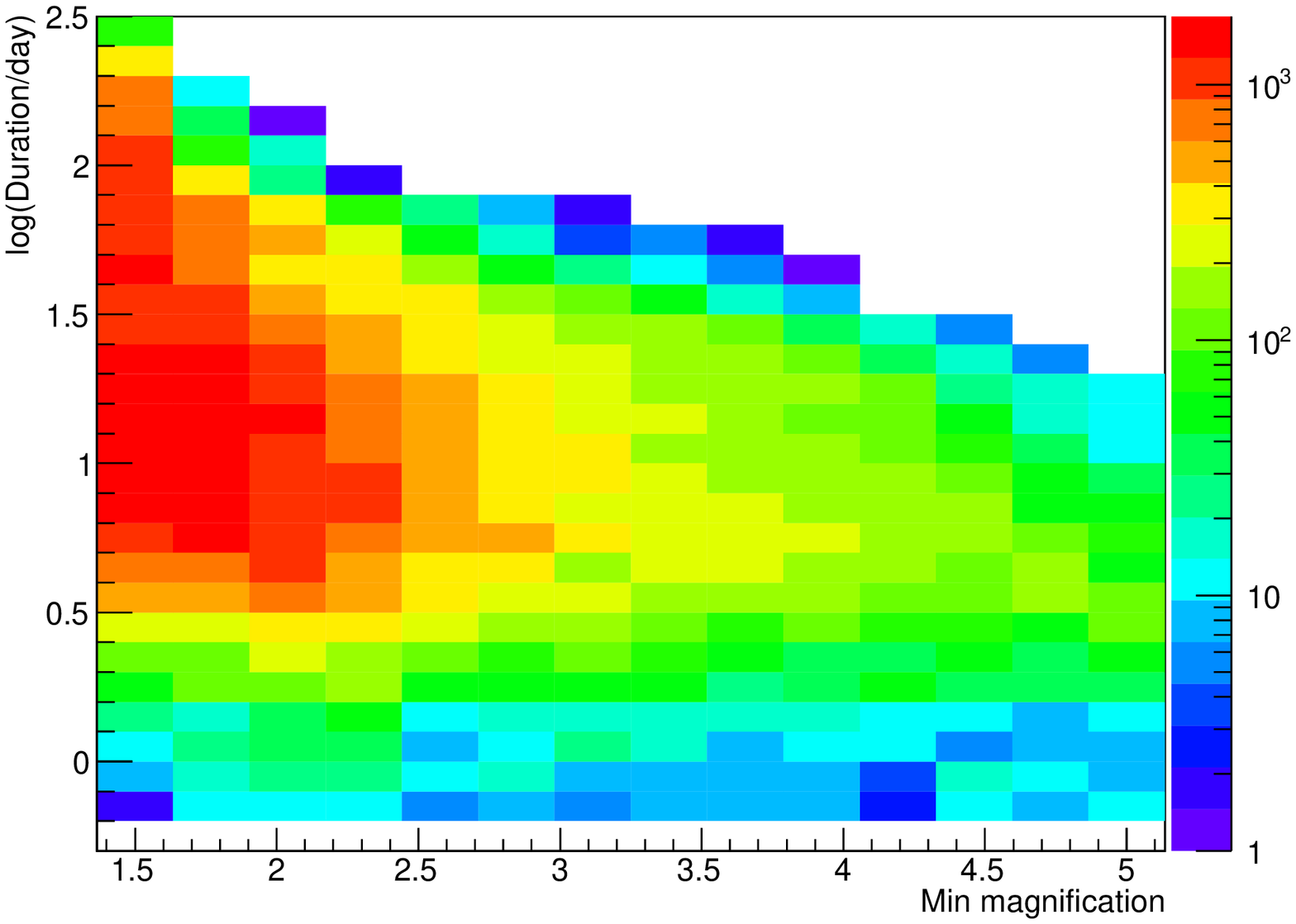}
\caption{
Time scales of different magnifications for a source crossing randomly selected paths in the inner 4$\times$4\,pc$^2$ region of GMC.
The source is moving with a superluminal speed of $5\,c$. 
The case of a flat distribution of clumps in GMC (using magnification map of \ref{fig_gmc1}) is shown in the top panel.
The case of a peaked distribution of clumps in GMC (using magnification map of \ref{fig_gmc2}) is shown in the bottom panel.
}\label{fig_gmc5}
\end{figure}

In order to study in detail the possible time scales for various magnifications, we simulated $10^4$ random paths through the central region of GMC. 
For each path we searched for the time periods during which a magnification was above a given value.
The results of the calculations, for a flat and a peaked distribution of clumps, are shown in Fig.~\ref{fig_gmc5}.
Due to the interplay between the influence of clumps with various sizes, the distribution of time scales for a flares of a given magnification is rather broad.
Also for some paths the emission might start in a region already magnified, therefore shortening the time scale compared with crossing through complete clump. 
The flares with a time scales of a few tens of days can be obtained for the peaked distribution of clumps up to a magnification factor of a few. 

\begin{figure}
\includegraphics[width=0.49\textwidth]{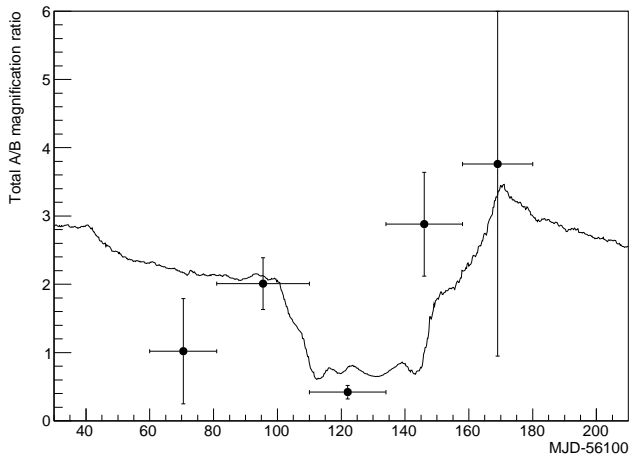}
\caption{
Changes of the total magnification ratio of the leading to trailing component seen in \textit{Fermi}-LAT data during 2012 high state of \srcs\ (black circles, \citealp{vn15}).
The black solid line shows the changes of the magnification ratio obtained along one of simulated paths of the trailing image through a GMC with mass of $2\times10^5M_\odot$.
Strong lensing with magnification ratio of 3.6, apparent superluminal speed of the source of $5\,c$ and size of the gamma-ray emission region of 0.01\,pc are assumed.
}\label{fig_fermi_magn}
\end{figure}

In order to explain the changes in the magnification ratio claimed by \citet{vn15} in the framework of this model, we compute the total magnification ratio as the product of the strong lensing magnification and the magnification from microlensing on clumps in a GMC. 
Based on the radio measurements we assume the value of the strong lensing magnification ratio of the leading to trailing image of $\sim3.6$ \citep{bi99}. 
If the trailing image is boosted by microlensing on clumps in a GMC, the total magnification ratio will decrease. 
In Fig.~\ref{fig_fermi_magn} we show one of possible paths obtained in the simulations which follows the changes of the magnification ratio seen in the \textit{Fermi}-LAT data.

One may question how stable is the magnification map being considered in the case of microlensing on the medium-sized structures. 
In particular, it is curious whether significant changes of the magnification pattern can happen during the 2012 and 2014 flaring period, assuming that the location of the emission zone is the same in both cases. 
The Keplerian \kom{velocity} of GC or GMC orbiting around the lensing galaxy at the distance $r$ is 
$v=670 (M/10^{11}M_\odot)^{1/2}(r/1\,\mathrm{kpc})^{-1/2}\,\mathrm{[km/s]}$, where $M$ is the part of the mass of the lensing galaxy contained within $r$.
The relative velocity of the source galaxy, lens, and the Milky Way is also expected to be of the same order (see the measurement of the Local Group with respect to the Cosmic Microwave Background, $627\pm22\,\mathrm{km\,s^{-1}}$, \citealp{ko93}). 
These speeds are about an order of magnitude larger than the movement of individual clumps in the GMC (see e.g. \citealp{sg90}). 
Therefore, we can expect shifts of the magnification patterns by $\sim 10^{-3}\,$pc per year, which is much smaller than the size of the individual clumps in a GMC.

On the other hand, the location of the emission region in \srcs\ might vary between different flaring states. 
Such possibility is further supported by clearly different GeV gamma-ray spectral shape in both cases \citep{bu15} and by a different type of activity (multiple flares in 2012 versus a single flare in 2014). 
In the case of flat spectrum radio quasars, the gamma-ray emission can be easily generated up to $R_\mathrm{em, max}=0.1$\,pc along the jet.
Assuming that the jet is visible at an angle of $\theta_\mathrm{jet}$, this corresponds to a distance (measured in the frame of the lens) of
$8\times 10^{-3} (R_\mathrm{em, max}/0.1\mathrm{pc}) (\theta_\mathrm{jet}/5^\circ)\mathrm{pc}$.
Therefore, in this scenario it is still expected that the microlensing magnification pattern does not change significantly for different flaring periods. 
Note however that one to two order of magnitude larger distances from the base of the jet are possible if the emission occurs as a comptonization of radiation of the dust torus rather than the broad line region.
In this case the changes of the magnification pattern on the yearly scale would be expected.
Therefore, the comparison of the evolution of the relative magnification ratio between different flaring periods might be another discriminant in the long standing problem of the location of the emission region in the FSRQs.
Finally, \kom{\citet{ch14} and} \citet{ba15} interpreted the difference in the time delay of the two images in the radio and in gamma rays as a difference in the projected distance of the radio core and the high energy emission region to be $\sim 50$\,pc.
If the location of the gamma-ray emission region can vary by a fraction of this number (which might be difficult taking into account the available radiation fields which might serve as a target for gamma-ray production), the projected location of the emission zone might move in and out of the region covered by a GMC.

One may wonder whether the occurrence of medium size structures like a GMC or a GC in the line of sight of one of the quasar images might affect the spectral shape of the observed gamma rays from those sources.
We note however \kom{that} the optical emission of stars in the GC is too weak to provide a strong absorption of the sub-TeV gamma-rays. 
On the other hand, the IR radiation from the GMCs would only affect TeV photons which are normally not observable from distant sources (such as lensed blazars) due to strong absorption by the extragalactic background light. 

\section{Conclusions}
We investigated the microlensing of \srcs\ on medium-size structures in its lensing galaxy \srclens . 
We studied the cases of GC, OC and GMC. 
We derived the probability of such an event and expected magnification values as well as their time scales.

We conclude that microlensing, occurring on clumps of a GMC, is a tempting alternative to explain the variability of the magnification factor seen in the GeV gamma-ray observations of \srcs .
Such a scenario is consistent with the current models of high-energy emission from blazars and, contrary to the microlensing on individual stars, is able to explain the high luminosity of a flare with simple relativistic boosting of the emission. 
The scenario is further supported by the measurements of large H$_2$ column densities in the direction of some of the images of the two lensed blazars in which short-term changes in the magnification due to microlensing were observed. 

\section*{Acknowledgments} 
We would like to thank Ievgen Vovk and the anonymous journal reviewer for their comments on the manuscript.
We would also like to thank John E. Ward for his English corrections of the manuscript text. 
This work is supported by the Polish NCN No. 2014/15/B/ST9/04043 grant.
JS is partially supported by Fundacja U\L .


\end{document}